\documentclass[11pt,a4paper]{article}
\pdfoutput=1
\usepackage{jheppub}
\usepackage{tabularx}
\usepackage{bibentry}    
\usepackage{amsmath}
\usepackage{amssymb}     
\usepackage{mathrsfs}    
\usepackage{mathtools}
\usepackage{dsfont}      
\usepackage{xcolor}      
\usepackage{xfrac}       
\usepackage{listings}

\definecolor{shadecolor}{rgb}{0.95,0.95,0.95}
\newcommand{\fpdi}{\mathrm{f}}
\newcommand{\Fpdi}{\mathrm{F}}

\renewcommand{\d}{\text{d}}

\newcommand{\eq}[1]{#1}

\newcommand{\eqsp}{\;}
\newcommand{\e}{\,}

\newcommand{\X}{x}

\newcommand{\ldefine}{\vcentcolon=}

\newcommand{\cmd}[1]{\vspace{0.3cm}\texttt{\$ #1}\\[0.3cm]}

\newcommand{\acropolis}{\texttt{ACROPOLIS}~}

\graphicspath{ {../plots/} }

\preprint{DESY 20-161, ULB-TH/20-16}

\title{\texttt{ACROPOLIS}: \texttt{A} generi\texttt{C} f\texttt{R}amework f\texttt{O}r \texttt{P}hotodisintegration \texttt{O}f \texttt{LI}ght element\texttt{S}}

\author{Paul Frederik Depta$^\text{a}$,}
\author{Marco Hufnagel$^{\text{a}, \text{b}}$, and}
\author{Kai Schmidt-Hoberg$^\text{a}$}

\affiliation{$^\text{a}$ DESY, Notkestra\ss e 85, D-22607 Hamburg, Germany}
\affiliation{$^\text{b}$ Service de Physique Théorique, Université Libre de Bruxelles, Boulevard du Triomphe, CP225, B-1050 Brussels, Belgium}

\emailAdd{frederik.depta@desy.de}
\emailAdd{marco.hufnagel@desy.de}
\emailAdd{kai.schmidt-hoberg@desy.de}

\abstract{
The remarkable agreement between observations of the primordial light element abundances and the corresponding theoretical predictions within the standard cosmological history provides a powerful method to constrain physics beyond the standard model of particle physics (BSM). For a given BSM model these primordial element abundances are generally determined by {\it (i) Big Bang Nucleosynthesis} and {\it (ii) possible subsequent disintegration processes}. The latter potentially change the abundances due to late-time high-energy injections which may be present in these scenarios. While there are a number of public codes for the first part, no such code is currently available for the second.
Here we close this gap and present \texttt{ACROPOLIS}, \texttt{A} generi\texttt{C} f\texttt{R}amework f\texttt{O}r \texttt{P}hotodisintegration \texttt{O}f \texttt{LI}ght element\texttt{S}. The widely discussed cases of decays as well as annihilations can be run without prior coding knowledge within example programs. Furthermore, due to its modular structure, \texttt{ACROPOLIS} can easily be extended also to other scenarios.
}

\begin{document}
\maketitle

\section{Introduction}

The remarkable overall agreement between the inferred abundances of light elements and the corresponding predictions within the standard model of particle physics (SM) implies that any deviation from
standard cosmology for sub-MeV temperatures is strongly constrained~\cite{Shvartsman:1969mm,Steigman:1977kc,Scherrer:1987rr,Cyburt:2015mya}. On the other hand, there is overwhelming evidence
for the existence of dark matter (DM) suggesting that physics beyond the SM (BSM) is necessary to explain all phenomena we observe in nature.
To evaluate the viability of a given BSM model it is therefore crucial to calculate its effect on the primordial element abundances and to compare them with the observationally inferred values.

There are two different temperature ranges in the early universe which determine the various element abundances:
Once the universe has cooled down to about 1~MeV -- at which point the typical energy of photons in the thermal bath has dropped significantly below the relevant binding energies --
protons and neutrons start to fuse into light nuclei such as deuterium, helium, and lithium. This process of {\it Big Bang Nucleosynthesis} (BBN) can be tracked
numerically by solving the appropriate Boltzmann equations describing the various fusion processes and a number of public codes such as \texttt{AlterBBN}~\cite{Arbey:2018zfh}, \texttt{PArthENoPE}~\cite{Pisanti:2007hk}, and \texttt{PRIMAT}~\cite{PitrouEtal2018} are available for this task. For most scenarios, the fusion processes have completed at about 10~keV and the nuclear abundances remain frozen at their asymptotic values until other processes become effective that might further change the abundances. For the standard cosmological history this would only happen at very late times, e.g.\ due to a reprocessing of the abundances in stellar fusion reactions.

However, many BSM models predict late-time decays or residual annihilations of dark sector particles which can lead to
subsequent nuclear disintegration processes due to electromagnetic or hadronic showers, further changing the `would-be' abundance values from BBN.
For rather {\it heavy} dark sectors both photodisintegration and hadrodisintegration are relevant and the treatment of photodisintegration is simplified by the fact that
the resulting photon spectrum has a universal form for sufficiently large injection energies depending only on the injection time and total energy injected~\cite{Cyburt:2002uv}.
Recently however the idea of {\it light} dark sectors with masses in the MeV to GeV range has attracted a lot of attention~\cite{Batell:2009di,Andreas:2012mt,Schmidt-Hoberg:2013hba,Essig:2013vha,Izaguirre:2013uxa,Batell:2014mga,Dolan:2014ska,Krnjaic:2015mbs,Dolan:2017osp,Izaguirre:2017bqb,Knapen:2017xzo,Beacham:2019nyx,Bondarenko:2019vrb,Filimonova:2019tuy}
and the light element abundances have been studied
for such setups~\cite{Hufnagel:2017dgo,Hufnagel:2018bjp,Forestell:2018txr,Depta:2019lbe,Depta:2020wmr,Kawasaki:2020qxm}.
What makes the correct treatment of such scenarios more involved is that the coupled evolution equations of all particles that can emerge in electromagnetic decays have to be solved explicitly,
as the approximation of the `universal photon spectrum' breaks down if the energy of the initial decay products is too low~\cite{Poulin:2015opa}.
However, unlike the case for BBN, there is currently no code available to treat possible late-time modifications of the abundances due to photodisintegration.
Here we close this gap and present \texttt{ACROPOLIS}, \texttt{A} generi\texttt{C} f\texttt{R}amework f\texttt{O}r \texttt{P}hotodisintegration \texttt{O}f \texttt{LI}ght element\texttt{S}.
The widely discussed cases of decays as well as annihilations are already implemented in example programs which can be run without prior coding knowledge.
Furthermore, due to its modular structure, \texttt{ACROPOLIS} can easily be extended also to other scenarios and could also be linked to larger computational frameworks such as
\texttt{CosmoBIT}~\cite{Renk:2020hbs}.

\section{Theoretical background}
We follow the procedure detailed in~\cite{Hufnagel:2018bjp}, which generalised the one from~\cite{Poulin:2015opa} by not only including photons but also electrons and positrons. To render this manual self-contained we repeat all relevant steps. Following the convention in the literature on photodisintegration, we consider the phase-space distribution function $\fpdi_x$ of particles $x \in \{e^\pm, \gamma \}$ differential in the energy $E$. Taking into account the degrees of freedom of $x$, $g_x$, this is related to the more conventional distribution $f_x$ differential in the momentum $p$ via
\begin{align}
\fpdi_x(E) = g_x f_x(p) \frac{E \, p}{2 \pi^2}\eqsp,
\end{align}
where $E = \sqrt{m_x^2 + p^2}$ is the energy with mass $m_x$ and $p$ is the momentum.

\subsection{Electromagnetic cascade}

The late-time injection of high-energetic electromagnetic particles into the SM plasma induces an electromagnetic cascade that leads to non-thermal parts of the photon, electron, and positron spectra. Denoting particles from the thermal background with a subscript `th' the most relevant interactions are
\begin{enumerate}
\item Double photon pair creation $\gamma \gamma_\mathrm{th} \rightarrow e^+ e^-$,
\item Photon-photon scattering $\gamma \gamma_\mathrm{th} \rightarrow \gamma \gamma$,
\item Bethe-Heitler pair creation $\gamma N \rightarrow e^+ e^- N$ with $N \in \{{}^1 \mathrm{H}, {}^4 \mathrm{He} \}$,
\item Compton scattering $\gamma e^-_\mathrm{th} \rightarrow \gamma e^-$, and
\item Inverse Compton scattering $e^\pm \gamma_\mathrm{th} \rightarrow e^\pm \gamma$.
\end{enumerate}
Other processes are suppressed by small number densities and can be neglected, cf.\ appendix~\ref{app:rates_cascade}. Before turning to the appropriate Boltzmann equation for the description of the electromagnetic cascade, let us briefly comment on why photodisintegration is only possible for late-time and high-energy injections into the SM plasma. For photons with energies above the threshold for double photon pair creation $E_{e^+ e^-}^\mathrm{th} \simeq m_e^2/(22 T)$~\cite{Kawasaki:1994sc}, where $m_e$ is the electron mass and $T$ is the photon temperature, this process is much more efficient than the other reactions, thus rapidly depleting these high-energy photons. This implies that any photodisintegration process can only occur if $E_{e^+ e^-}^\mathrm{th}$ is above the threshold for the various disintegration reactions, i.e.\ if $T$ is small enough translating to
\begin{enumerate}
\item $T \lesssim 5.34 \, \mathrm{keV}$ for $\mathrm{D}$-disintegration with $E_\mathrm{D}^\mathrm{th} \approx 2.22 \, \mathrm{MeV}$,
\item $T \lesssim 1.90 \, \mathrm{keV}$ for ${}^3\mathrm{H}$-disintegration with $E_{{}^3\mathrm{H}}^\mathrm{th} \approx 6.26 \, \mathrm{MeV}$,
\item $T \lesssim 2.16 \, \mathrm{keV}$ for ${}^3\mathrm{He}$-disintegration with $E_{{}^3\mathrm{He}}^\mathrm{th} \approx 5.49 \, \mathrm{MeV}$,
\item $T \lesssim 0.60 \, \mathrm{keV}$ for ${}^4\mathrm{He}$-disintegration with $E_{{}^4\mathrm{He}}^\mathrm{th} \approx 19.81 \, \mathrm{MeV}$,
\item $T \lesssim 3.21 \, \mathrm{keV}$ for ${}^6\mathrm{Li}$-disintegration with $E_{{}^6\mathrm{Li}}^\mathrm{th} \approx 3.70 \, \mathrm{MeV}$,
\item $T \lesssim 4.81 \, \mathrm{keV}$ for ${}^7\mathrm{Li}$-disintegration with $E_{{}^7\mathrm{Li}}^\mathrm{th} \approx 2.47 \, \mathrm{MeV}$, and
\item $T \lesssim 7.48 \, \mathrm{keV}$ for ${}^7\mathrm{Be}$-disintegration with $E_{{}^7\mathrm{Be}}^\mathrm{th} \approx 1.59 \, \mathrm{MeV}$.
\end{enumerate}
We observe that photodisintegration will become effective for rather low temperatures where BBN has already finished, so that the two processes simply factorise.
This implies in particular that the initial abundances for photodisintegration correspond to the final abundances of BBN.
Even though we do not employ it within \texttt{ACROPOLIS}, let us also remark that for injection energies above $E_{e^+ e^-}^\mathrm{th}$, the non-thermal part of the photon spectrum is close to the universal spectrum~\cite{Cyburt:2002uv}
\begin{align}
\mathrm{f}_{\gamma, \mathrm{univ}} (E) \sim \begin{cases} K_0 (E/E_X)^{-3/2} &\text{for} \; E < E_X\eqsp, \\
K_0 (E/E_X)^{-2} &\text{for} \; E_X < E < E_{e^+ e^-}^\mathrm{th}\eqsp, \\
0 &\text{for} \; E > E_{e^+ e^-}^\mathrm{th}\eqsp, \end{cases}
\end{align}
where $K_0 = E_0 E_X^{-2} [2 + \ln (E_{e^+ e^-}^\mathrm{th} / E_X)]^{-1}$ and $E_X = m_e^2 / (80 T)$. In \texttt{ACROPOLIS} we solve the full Boltzmann equation numerically in which case
the cutoff at $E_{e^+ e^-}^\mathrm{th}$ is replaced by an exponential suppression
reflecting the aforementioned suppression of photodisintegration reactions due to efficient double photon pair creation.

As the total interaction rates $\Gamma_x$ are large compared to the Hubble rate $H$, the expansion of the universe can be neglected in the calculation of the non-thermal photon spectrum~\cite{Cyburt:2002uv,Jedamzik:2006xz}. Suppressing the $t$- and $T(t)$-dependencies, the Boltzmann equation reads
\begin{align}
\frac{\partial \fpdi_\X(E)}{\partial t} \simeq S_\X(E) - \Gamma_\X(E)\fpdi_\X(E) + \sum_{\X'}\int_E^\infty \d E' \, K_{\X' \rightarrow \X}(E, E')\fpdi_{\X'}(E')\eqsp,
\label{eq:boltzmann_cascade}
\end{align}
where $S_x (E)$ is the source term for the production of $x$ with energy $E$, $\Gamma_x (E)$ is the total interaction rate for $x$ with energy $E$, and $K_{x' \rightarrow x} (E, E')$ is the differential interaction rate going from particle $x'$ with energy $E'$ to particle $x$ with energy $E$. These rates contain the electromagnetic cascade reactions listed above and are given in appendix~\ref{app:rates_cascade}. The electromagnetic cascade reactions quickly establish a quasi-static equilibrium with $\partial \fpdi_x / \partial t \simeq 0$~\cite{Cyburt:2002uv,Jedamzik:2006xz}, hence we search for solutions of the integral equation
\begin{align}
\fpdi_\X(E) = \frac{1}{\Gamma_\X(E)} \left( S_\X(E) + \sum_{\X'} \int_E^\infty K_{\X' \rightarrow \X}(E, E')\fpdi_{\X'}(E')\,\d E' \right)\eqsp.
\label{eq:cascade_f}
\end{align}

We consider source terms arising from monochromatic high-energy injections, as typically realised for decays or residual annihilations of non-relativistic particles into two-particle SM final states.
These can be parametrised via
\begin{align}
S_\X(E) = S_\X^{(0)} \delta(E-E_0) + S_\X^{(\text{FSR})}(E)\eqsp,
\label{eq:SXE_definition}
\end{align}
where $E_0$ is the injection energy and the final-state radiation part $S_\X^{(\text{FSR})}$ is zero for electron-positron pairs ($S_{e^\pm}^{(\text{FSR})} \equiv 0$) and proportional to the monochromatic injection for photons ($S_{\gamma}^{(\text{FSR})} \propto S_{e^\pm}^{(0)}$).

For the numerical solution of eq.~\eqref{eq:cascade_f} we subtract the term containing the $\delta$-distribution by defining
\begin{align}
\Fpdi_\X(E) \ldefine \fpdi_\X(E) - \frac{S_\X^{(0)} \delta(E-E_0)}{\Gamma_\X(E)}
\label{eq:Ff_relation}
\end{align}
such that
\begin{align}
\Gamma_\X(E) \Fpdi_\X(E) = S_{\X}^{(\text{FSR})}(E) + \sum_{\X'} \left[ \frac{K_{\X' \to \X} (E, E_0) S_{\X'}^{(0)}}{\Gamma_{\X'} (E_0)} + \int_{E}^{\infty} K_{\X' \to \X} (E, E') \Fpdi_{\X'}(E')\,\d E'\right] \eqsp.
\label{eq:cascade_full}
\end{align}

\subsection{Non-thermal nucleosynthesis}

With the previously calculated photon spectra $\fpdi_\gamma$ we can now evaluate the effect on the primordial light element abundances $N \in \{n, p, D, {}^3\mathrm{H}, {}^3\mathrm{He}, {}^4\mathrm{He}, {}^6\mathrm{Li}, {}^7\mathrm{Li}, {}^7\mathrm{Be} \}$. The corresponding Boltzmann equation is given by
\begin{align}
\dot{Y}_N(t) = \sum_{j} Y_{j}(t) \int_{0}^{\infty} \text{d} E \, \fpdi_\gamma(t, E)\sigma_{j\gamma \rightarrow N}(E) - Y_N(t) \sum_{j'} \int_{0}^{\infty} \text{d} E \, \fpdi_\gamma(t, E)\sigma_{N\gamma \rightarrow j'}(E)\eqsp,
\label{eq:y_pdi}
\end{align}
where $Y_N = n_N / n_b$ with $n_N$ $(n_b)$ the number density of $N$ (baryons) and $\sigma_r$ the cross section for the reaction $r$. Here we implement all reactions that are shown in tab.~\ref{tab:reactions}
\begin{table}
	\centering
	\begin{tabular}{|rcrcrcrcr|c|}
		\hline
		& & & & & & & & & $E^\text{th}\;\;[\mathrm{MeV}]$\\
		\hline
		$\mathrm{D}$ & $+$ & $\gamma$ & $\rightarrow$ & $p$ & $+$ & $n$ & & & $2.22$\\
		\hline
		$^3\mathrm{H}$ & $+$ & $\gamma$ & $\rightarrow$ & $\mathrm{D}$ & $+$ & $n$ & & & $6.26$\\
		\hline
		$^3\mathrm{H}$ & $+$ & $\gamma$ & $\rightarrow$ & $p$ & $+$ & $n$ & $+$ & $n$ & $8.48$ \\
		\hline
		$^3\mathrm{He}$ & $+$ & $\gamma$ & $\rightarrow$ & $\mathrm{D}$ & $+$ & $p$ & & & $5.49$ \\
		\hline
		$^3\mathrm{He}$ & $+$ & $\gamma$ & $\rightarrow$ & $n$ & $+$ & $p$ & $+$ & $p$ & $7.12$\\
		\hline
		$^4\mathrm{He}$ & $+$ & $\gamma$ & $\rightarrow$ & $^3\mathrm{H}$ & $+$ & $\, p$ & & & $19.81$ \\
		\hline
		$^4\mathrm{He}$ & $+$ & $\gamma$ & $\rightarrow$ & $^{3}\mathrm{He}$ & $+$ & $n$ & & & $20.58$ \\
		\hline
		$^4\mathrm{He}$ & $+$ & $\gamma$ & $\rightarrow$ & $\mathrm{D}$ & $+$ & $\mathrm{D}$ & & & $23.84$ \\
		\hline
		$^4\mathrm{He}$ & $+$ & $\gamma$ & $\rightarrow$ & $\mathrm{D}$ & $+$ & $n$ & $+$ & $p$ & $26.07$ \\
		\hline
		$^6\mathrm{Li}$ & $+$ & $\gamma$ & $\rightarrow$ & $\, ^{4}\mathrm{He}$ & $+$ & $n$ & $+$ & $p$ & $3.70$ \\
		\hline
		$^6\mathrm{Li}$ & $+$ & $\gamma$ & $\rightarrow$ & $\mathrm{X}$ & $+$ &  $\, ^{3}A$ & & & $15.79$ \\
		\hline
		$^7\mathrm{Li}$ & $+$ & $\gamma$ & $\rightarrow$ & $^3\text{H}$ & $+$ &  $\, ^{4}\text{He}$ & & & $2.47$ \\
		\hline
		$^7\mathrm{Li}$ & $+$ & $\gamma$ & $\rightarrow$ & $n$ & $+$ &  $\, ^{6}\text{Li}$ & & & $7.25$ \\
		\hline
		$^7\mathrm{Li}$ & $+$ & $\gamma$ & $\rightarrow$ & $2n$ & $+$ &  $p$ & $+$ & $ ^{4}\text{He}$ & $10.95$ \\
		\hline
		$^7\mathrm{Be}$ & $+$ & $\gamma$ & $\rightarrow$ & $^3\text{He}$ & $+$ &  $ ^{4}\text{He}$ & & & $1.59$ \\
		\hline
		$^7\mathrm{Be}$ & $+$ & $\gamma$ & $\rightarrow$ & $p$ & $+$ &  $ ^{6}\text{Li}$ & & & $5.61$ \\
		\hline
		$^7\mathrm{Be}$ & $+$ & $\gamma$ & $\rightarrow$ & $2p$ & $+$ &  $n$ & $+$ & $ ^{4}\text{He}$ & $9.30$ \\
		\hline
	\end{tabular}
\caption{Reactions and threshold energies from~\cite{Cyburt:2002uv} for the processes we consider in eq.~\eqref{eq:y_pdi}.}
\label{tab:reactions}
\end{table}
by adopting the analytical expressions for the rates $1-17$ from~\cite{Cyburt:2002uv}; however, we modify the prefactor of reaction $7$ from $17.1\e\mathrm{mb}$ to $20.7\e\mathrm{mb}$ as suggested by~\cite{Jedamzik:2006xz} in order to match the most recent EXFOR data. Note that these disintegration reactions can be neglected in eq.~\eqref{eq:boltzmann_cascade} due to the low number density of nuclei, thus enabling us to calculate the photon spectrum without knowledge of the light element abundances.

\section{Numerical solution techniques}

\subsection{Electromagnetic cascade}
For the numerical solution of eq.~\eqref{eq:cascade_full} we can exploit the fact that only particles with energies less than the injection energy can be produced, leading to vanishing spectra for $E > E_0$. Moreover, we are only interested in energies above a minimal energy $E_\text{min}$ given by the lowest threshold energy of the photodisintegration reactions. In the code we therefore set $E_\text{min} = 1.5\e\mathrm{MeV}$ ($E_{{}^7\mathrm{Be}}^\mathrm{th} \approx 1.59 \, \mathrm{MeV}$). With the relevant energy range $[E_\text{min}, E_0]$ we then define a grid of energies spaced evenly on a log-scale, $\epsilon_i \ldefine E_\text{min} \times (E_0 / E_\text{min})^{i/(M-1)}$, where $i \in \left\{ 0, \dots, M-1 \right\}$, $(\epsilon_0, \epsilon_{M-1}) = (E_\text{min}, E_0)$. By default we choose $150$ points per decade, which is usually sufficient to ensure convergence for all parameter points. Evaluating eq.~\eqref{eq:cascade_full} at each individual grid point we then find~\cite{Hufnagel:2020nxa}
\begin{align}
\Gamma_\X(\epsilon_i) \Fpdi_\X(\epsilon_i)  &= S_\X^{\text{(FSR)}}(\epsilon_i) + \sum_{\X'} \left[ \frac{K_{\X' \to \X} (\epsilon_i, E_0) S_{\X'}^{(0)}}{\Gamma_{\X'} (E_0)} + \int_{\ln(\epsilon_i)}^{\ln(E_0)} \text{d} y \, e^y K_{\X' \to \X} (\epsilon_i, e^y) \Fpdi_{\X'}(e^y)\right] \nonumber \\[4mm]
&\simeq S_\X^{\text{(FSR)}}(\epsilon_i) +  \sum_{\X'} \bigg[ \frac{K_{\X' \to X} (\epsilon_i, E_0) S_{\X'}^{(0)}}{\Gamma_{X'} (E_0)} + \frac{\Delta y}{2} \bigg( 2 \sum_{j=i+1}^{M-2} \epsilon_j K_{\X' \to \X} (\epsilon_i, \epsilon_j) \Fpdi_{\X'}(\epsilon_j)\nonumber\\
&\qquad\quad\qquad\qquad\quad\;\;\; +\epsilon_i K_{\X' \to \X} (\epsilon_i, \epsilon_i) \Fpdi_{\X'}(\epsilon_i)\,  +\,  E_0 K_{\X' \to \X} (\epsilon_i, E_0) \Fpdi_{\X'}(E_0) \bigg) \bigg]\eqsp,
\label{eq:FXi}
\end{align}
with $\Delta y = \ln(E_0/E_\text{min})/(M-1)$. In the last step, we have used the trapezoidal integration rule and the sum $\sum_{j=i+1}^{M-2}$ is understood to vanish for $i +1 > M - 2$. This expression is valid for $i < M -1$, while for $i=M-1$ we simply have
\begin{align}
\Fpdi_\X(E_0) = \frac{S_\X^{\text{(FSR)}}(E_0)}{\Gamma_\X(E_0)} + \sum_{\X'} \frac{K_{\X' \to \X} (E_0, E_0) S_{\X'}^{(0)}}{\Gamma_\X(E_0) \Gamma_{\X'} (E_0)}\eqsp.
\label{eq:FX_ES}
\end{align}
Assuming that $\Fpdi_\X(E_j)$ has already been calculated for $j>i$, eq.~(\ref{eq:FXi}) can be interpreted as a linear system of three equations for the unknown variables $\Fpdi_\X(E_i)$. Consequently, by defining $\bar{\Fpdi}(\epsilon_i) \ldefine \big[ \Fpdi_\gamma(\epsilon_i), \Fpdi_{e^-}(\epsilon_i), \Fpdi_{e^+}(\epsilon_i)\big]^\text{T}$, we have
\begin{align}
\bar{\Fpdi}(\epsilon_i) = a(\epsilon_i) + B(\epsilon_i) \bar{\Fpdi}(\epsilon_i)
\label{eq:Fab}
\end{align}
with
\begin{align}
\big[ a(\epsilon_i) \big]_{\X\hphantom{\X'}} & \ldefine \frac{1}{\Gamma_\X(\epsilon_i)} \sum_{\X'} \bigg[ \frac{K_{\X' \to \X} (\epsilon_i, E_0) S_{\X'}^{(0)}}{\Gamma_{\X'} (E_0)} + \frac{\Delta y}{2} \bigg( 2 \sum_{j=i+1}^{M-2} \epsilon_j K_{\X' \to \X} (\epsilon_i, \epsilon_j) F_{\X'}(\epsilon_j)\nonumber\\
& \hspace{4cm} + E_0 K_{\X' \to \X} (\epsilon_i, E_0) F_{\X'}(E_0) \bigg) \bigg] + \frac{S_\X^{\text{(FSR)}}(\epsilon_i)}{\Gamma_\X(\epsilon_i)}\eqsp, \\[4mm]
\big[B(\epsilon_i)\big]_{\X \X'} & \ldefine \frac{\Delta y}{2} \frac{\epsilon_i K_{\X' \to \X} (\epsilon_i, \epsilon_i)}{\Gamma_\X(\epsilon_i)}\eqsp.
\end{align}
Given the knowledge of $\big[ a(\epsilon_i) \big]_{\X}$ and $\big[B(\epsilon_i)\big]_{\X \X'}$, the linear eq.~\eqref{eq:Fab} can then be solved using standard techniques to calculate the values of $\bar{\Fpdi}(\epsilon_i)$.
As $\big[ a(\epsilon_i) \big]_{\X}$ and $\big[B(\epsilon_i)\big]_{\X \X'}$ explicitly depend on $\bar{\Fpdi}(\epsilon_j)$ for $j > i$, we start solving eq.~\eqref{eq:FX_ES} at $i=M-1$ to successively determine solutions of eq.~\eqref{eq:Fab} at $i < M-1$.

\subsection{Non-thermal nucleosynthesis}
For the numerical solution of eq.~\eqref{eq:y_pdi} we first define
\begin{align}
\Gamma_r(t) \ldefine \int_0^\infty \fpdi_\gamma(t, E) \sigma_r(E)\,\d E\eqsp,
\end{align}
which transforms eq.~\eqref{eq:y_pdi} into the form
\begin{align}
\dot{Y}_N(t) = \sum_{j} Y_{j}(t) \Gamma_{j\gamma \rightarrow N}(t) - Y_N(t) \sum_{j'} \Gamma_{N\gamma \rightarrow j'}(t)\eqsp.
\label{eq:y_pdi_redef}
\end{align}
Consequently, by also defining $\bar{Y}(t) \ldefine \big[ Y_n(t), Y_p(t), Y_\text{D}(t), ... \big]^T$ and after substituting $t \rightarrow T$ by means of the time-temperature relation, we find
\begin{equation}
\frac{\d \bar{Y}(T)}{\d T} = \mathcal{R}(T) \bar{Y}(T)
\label{eq:YT}
\end{equation}
with
\begin{align}
\big[ \mathcal{R}(T)\big]_{N N'} \ldefine \frac{\d t}{\d T} \times \left[ \Gamma_{N'\gamma\rightarrow N}(T)  - \delta_{N N'} \sum_{j'} \Gamma_{N\gamma\rightarrow j'}(T)\right]\eqsp.
\label{eq:rate_matrix}
\end{align}
Here, the matrix $\big[ \mathcal{R}(T)\big]_{N N'}$ can be obtained via numerical integration, and eq.~\eqref{eq:YT} is an ordinary system of differential equations, which is solved by
\begin{align}
\bar{Y}(T) = \exp\left(\int_{T_\text{max}}^T \mathcal{R}(T')\,\d T'\right)\bar{Y}_0
\end{align}
with the initial condition $\bar{Y}(T_\text{max}) = \bar{Y}_0$ at some maximal (initial) temperature $T_\text{max}$. The matrix exponential $\exp(\cdot)$ can be evaluated numerically by \emph{(i)} diagonalising the matrix $\int_{T_0}^T \mathcal{R}(T')\d T'$ with the corresponding unitary transformation $U_{\mathcal{R}}(T)$, \emph{(ii)} taking the exponential of the diagonal matrix by exponentiating each eigenvalue individually, and \emph{(iii)} transforming the resulting matrix back using $U_{\mathcal{R}}^{-1}(T)$.

\section{Example models}
\label{sec:example_models}

In this section, we briefly describe the physics of the two example models which are implemented in \texttt{ACROPOLIS}. In the next section, we will then discuss how to utilise the corresponding python scripts
\texttt{decay} and \texttt{annihilation}.

\subsection{Decay of a decoupled MeV-scale BSM particle}
\label{sec:decay_model}
As a first example model implemented in \texttt{ACROPOLIS}, we consider the non-relativistic decay of a decoupled $\mathrm{MeV}$-scale BSM particle $\phi$ with mass $m_\phi$ and lifetime $\tau_\phi$, e.g.\ an $\mathrm{MeV}$-scale mediator. This implementation closely follows~\cite{Hufnagel:2018bjp,em-decay}, but only considers a number density of $\phi$ fixed at a reference temperature $T_0$ that simply redshifts, i.e.\ remains comovingly constant, until it decays with a lifetime $\tau_\phi$. As pointed out in \cite{em-decay} this is in fact a consistent assumption for the parameter ranges where photodisintegration may be relevant, even when taking into account inverse decays of $\phi$.

The source terms entering in eq.~\eqref{eq:SXE_definition} for the non-relativistic decay of a decoupled BSM particle $\phi$ with mass $m_\phi$ and lifetime $\tau_\phi$ is given by~\cite{Hufnagel:2018bjp,Forestell:2018txr,Mardon:2009rc,Birkedal:2005ep}
\begin{align}
E_0 &= \frac{m_\phi}{2}\eqsp, \\
S_\gamma^{(0)} &= \text{BR}_{\gamma \gamma} \times \frac{2 n_\phi}{\tau_\phi}\eqsp,\label{eq:sa_decay} \\
S_{e^-}^{(0)} &= S_{e^+}^{(0)} = \text{BR}_{e^+ e^-} \times \frac{n_\phi}{\tau_\phi}\eqsp,\label{eq:se_decay} \\
S_\gamma^{(\text{FSR})} (E) &= \frac{S^{(0)}_{e^\pm}}{E_0} \times \frac{\alpha}{\pi} \frac{1 + (1-x)^2}{x} \ln \left( \frac{4 E_0^2 (1-x)}{m_e^2} \right) \times \Theta \left( 1 - \frac{m_e^2}{4 E_0^2} - x \right)\eqsp, \label{eq:safsr_decay}\\
S_{e^-}^{(\text{FSR})} (E) &= S_{e^+}^{(\text{FSR})} (E) = 0\eqsp,
\end{align}
where $n_\phi$ is the number density of $\phi$, $\text{BR}_{\gamma \gamma}$ ($\text{BR}_{e^+ e^-}$) is the branching ratio for decays into two photons (electron-positron pairs), $\alpha$ is the fine-structure constant, $m_e$ is the electron mass, and $x = E/E_0$. If there is no contribution from inverse decays to the abundance of $\phi$ below a reference temperature $T_0$, i.e.\ for $T_0 \ll m_\phi$, with corresponding time $t_0$, its number density can be parametrised as ($t \geq t_0$)
\begin{align}
n_\phi (t) \simeq n_\phi (T_0) \times \left( \frac{R(t_0)}{R(t)} \right)^3 \exp \left( - \frac{t}{\tau_\phi} \right)\eqsp,
\label{eq:n_decay}
\end{align}
where $R$ is the scale factor.

Whenever relevant, photodisintegration provides a very powerful probe of decaying MeV-scale BSM particles, constraining even very small abundances. We can thus assume that the energy density in the universe is dominated by the SM (and at late times also DM) energy densities, i.e.\ $m_\phi n_\phi \ll \rho_\mathrm{SM} + \rho_\mathrm{DM}$, and that the time-temperature relation of the SM is not changed by any BSM processes for $t > t_0$. This enables us to provide tables for the SM (photon) temperature, its derivative w.r.t.\ time, the neutrino temperature, the Hubble rate, and the scale factor.
Further approximating that standard BBN is not changed by any BSM processes including the presence and decay of $\phi$ implies that no separate BBN calculation must be performed and the results of a calculation using \texttt{AlterBBN} v1.4~\cite{Arbey:2011nf,Arbey:2018zfh} are provided.
The photodisintegration calculation is performed for temperatures $T_\mathrm{max} \geq T \geq T_\mathrm{min}$ with
\begin{align}
T_\mathrm{max} &= T(t = \tau_\phi) \times 10^{1/2} \eqsp, \\
T_\mathrm{min} &= T(t = \tau_\phi) \times 10^{-3/2}
\end{align}
around the decay time giving results of very good accuracy.

With the python script \texttt{decay} we thus provide an example program to calculate the primordial light element abundances after photodisintegration due to the decay of a BSM particle $\phi$ with mass $m_\phi$, lifetime $\tau_\phi$, a reference temperature $T_0$ with corresponding number density $n_\phi (T_0)$, and branching ratio $\text{BR}_{e^+ e^-}$ ($\text{BR}_{\gamma \gamma}$) for decays in electron-positron pairs (two photons).

\subsection{Residual annihilations of MeV-scale dark matter}
\label{sec:annihilation_model}

For the second example model implemented in \texttt{ACROPOLIS}, we consider residual annihilations of DM (or more generally any annihilating BSM particle) following~\cite{Depta:2019lbe}. We assume that DM consists of a self-conjugate\footnote{While only the case for self-conjugate DM is directly implemented in \texttt{ACROPOLIS}, the case of non self-conjugate DM differs only by a factor of 2 in $\langle \sigma v \rangle$. This factor is exact, unlike for freeze-out calculations~\cite{Bringmann:2020mgx}.} fermion $\chi$ with mass $m_\chi$ and an abundance (number density $n_\chi$) fixed such that $\Omega_\mathrm{DM} h^2 = 0.12$~\cite{Aghanim:2018eyx}.
The source terms entering in eq.~\eqref{eq:SXE_definition} are then given by~\cite{Depta:2019lbe}
\begin{align}
E_0 &= m_\chi \eqsp, \\
S_\gamma^{(0)} &= \text{BR}_{\gamma \gamma} \times \langle \sigma v \rangle n_\chi^2\eqsp, \\
S_{e^-}^{(0)} &= S_{e^+}^{(0)} = \text{BR}_{e^+ e^-} \times \frac{1}{2} \langle \sigma v \rangle n_\chi^2 \eqsp, \\
S_\gamma^{(\text{FSR})} (E) &=  \frac{S^{(0)}_{e^\pm}}{E_0} \times \frac{\alpha}{\pi} \frac{1 + (1-x)^2}{x} \ln \left( \frac{4 E_0^2 (1-x)}{m_e^2} \right) \times \Theta \left( 1 - \frac{m_e^2}{4 E_0^2} - x \right)\eqsp, \\
S_{e^-}^{(\text{FSR})} (E) &= S_{e^+}^{(\text{FSR})} (E) = 0\eqsp,
\end{align}
where $\text{BR}_{\gamma \gamma}$ ($\text{BR}_{e^+ e^-}$) is the branching ratio for annihilations into two photons (electron-positron pairs), $\alpha$ is the fine-structure constant, $m_e$ is the electron mass, $x = E/E_0$, and we expand the thermally averaged annihilation cross section\footnote{This is the sum of the one for annihilations in two photons and in electron-positron pairs, weighted according to their branching ratios.} in powers of the relative velocity $v_\mathrm{rel}$
\begin{align}
\langle \sigma v \rangle \simeq a + b \langle v_\mathrm{rel}^2 \rangle\eqsp.
\end{align}
Here, $s$-wave DM annihilations are dominated by $a$ whereas $p$-wave DM annihilations are dominated by $b$. The thermally averaged relative velocity squared is given by
\begin{align}
\langle v_\mathrm{rel}^2 \rangle \simeq \frac{6 T_\chi (T)}{m_\chi}\eqsp,
\end{align}
where $T_\chi$ is the DM temperature evolving before and after kinetic decoupling from the photon heat bath at $T^\mathrm{kd}$ according to
\begin{align}
T_\chi (T) = \begin{cases} T & \text{if}~T \geq T^\mathrm{kd} \eqsp, \\
T^\mathrm{kd} (R(T^\mathrm{kd}) / R(T))^2 & \text{if}~T < T^\mathrm{kd}\eqsp. \end{cases}
\end{align}
Similar to section~\ref{sec:decay_model} we assume that the energy density in the universe is dominated by the SM and DM $\chi$, and that the time-temperature relation of the SM is not influenced by the presence of $\chi$ or any other BSM particle for the temperatures relevant for photodisintegration. Also assuming standard BBN without any alterations due to BSM processes enables us to use the same data as before for the light element abundances as well as all temperatures, the Hubble rate, and the scale factor. The calculation is performed for temperatures $T_\mathrm{max} \geq T \geq T_\mathrm{min}$ with
\begin{align}
T_\mathrm{max} &= 2 \times \frac{m_e^2}{22 E_\mathrm{min}} \eqsp, \\
T_\mathrm{min} &= T_\mathrm{max} \times 10^{-4}\eqsp,
\end{align}
where $E_\mathrm{min} = 1.5 \, \mathrm{MeV}$ is just below the lowest disintegration threshold ($E_{{}^7\mathrm{Be}}^\mathrm{th} \approx 1.59 \, \mathrm{MeV}$). This range gives results of very good accuracy, as for larger temperatures the non-thermal photon spectrum is exponentially suppressed for energies above any disintegration threshold, while for smaller temperatures the annihilation rate is suppressed at least $\propto T^{-6}$.

To summarise, we provide the example program \texttt{annihilation} to calculate the primordial light element abundances after photodisintegration due to residual annihilations of DM particles $\chi$ with mass $m_\chi$, $s$ ($p$)-wave part $a$ ($b$) of the thermally averaged annihilation cross section, kinetic decoupling temperature $T^\mathrm{kd}$, and branching ratio into electron-positron pairs $\text{BR}_{e^+ e^-}$ (two photons $\text{BR}_{\gamma \gamma}$).

\section{Running \texttt{ACROPOLIS}}

\subsection{Installation}
\acropolis is open-source software, licensed under GPL3. The source code is publicly available and can be cloned from GitHub by executing the command\\\\
\cmd{git clone https://github.com/skumblex/acropolis.git}
which creates a new folder named \texttt{acropolis} with the repository inside of your current working directory.
\acropolis has been tested with \texttt{Python} version $\geq 3.7$ on macOS 10.15 and Ubuntu 20.04, and the following packages must be available (older versions might also work but have not been thoroughly tested):
\begin{itemize}
	\item \href{https://numpy.org/}{\texttt{NumPy}} $\geq 1.19.1$ (matrix manipulation)
	\item \href{https://www.scipy.org/}{\texttt{SciPy}} $\geq 1.5.2$ (numerical integration)
	\item \href{https://numba.pydata.org/}{\texttt{Numba}} $\geq 0.51.2$ (just-in-time compilation)
\end{itemize}
The most recent versions of these dependencies can be collectively installed at user-level, i.e. without the need for root access, by executing the command\\\\
\cmd{python3 -m pip install numpy, scipy, numba --user}
If these dependencies conflict with those for other programs in your work environment, it is strongly advised to utilise the capabilities of Python's virtual environments.

Optionally, pre-generated database files for the different (differential) reaction rates are available, whose presence significantly reduces the runtime of \acropolis for most parameter points, albeit at the cost of a somewhat higher RAM usage.\footnote{Several benchmark points that show the runtime of the code with and without the additional database files, can be found in appendix~\ref{sec:benchmarks_single}.} To download the respective database files with a total size of $\sim 490 \, \mathrm{MB}$, simply run the command\\\\
\cmd{./download\_db}
from within the \texttt{acropolis} directory, i.e.\ the one that was created while cloning the respective source code from GitHub.

\subsection{Using the predefined models}
\acropolis is designed to allow for an easy implementation of arbitrary models, but also contains reference implementations for the example models that were introduced in section~\ref{sec:example_models}. In fact, these two models and the corresponding scripts that are bundled with \acropolis should already suffice to treat most conceivable scenarios, implying that -- for most applications -- no additional coding is required. In the present section we will therefore first discuss how to correctly utilise these reference models, and only afterwards discuss how to implement additional ones in section~\ref{sec:implementing_models}.

\subsubsection{Running the wrapper scripts}
Within the \acropolis main directory there are two executables, \texttt{decay} and \texttt{annihilation}, which wrap the scenarios discussed in section~\ref{sec:decay_model} and section~\ref{sec:annihilation_model}, respectively. Both of these files need to be called with six command-line arguments each, a list of which can be obtained by running the command of choice without any arguments at all. Using the notation from section~\ref{sec:example_models}, the required parameters are
\begin{itemize}
	\item \texttt{decay}: $\hspace{1.8cm} m_\phi\;\,[\mathrm{MeV}] \quad \tau_\phi\;\,[\mathrm{s}] \quad T_0\;\,[\mathrm{MeV}] \quad n_\phi/n_\gamma|_{T_0} \quad \text{BR}_{ee} \quad \text{BR}_{\gamma\gamma}$
	\item \texttt{annihilation}: $\quad m_\chi\;\,[\mathrm{MeV}] \quad a\;\,[\mathrm{cm^3/s}] \quad b\;\,[\mathrm{cm^3/s}] \quad T_\text{kd}\;\,[\mathrm{MeV}] \quad \text{BR}_{ee} \quad \text{BR}_{\gamma\gamma}$
\end{itemize}
For example, the command\\[4mm]
\cmd{./decay 10 1e5 10 1e-10 0 1}
calculates the abundances after photodisintegration in the presence of an unstable particle $\phi$ with mass $m_\phi=10\,\mathrm{MeV}$, lifetime $\tau_\phi=10^5\,\mathrm{s}$, and number density $ n_\phi/n_\gamma|_{T_0} = 10^{-10}$ at $T_0=10\,\mathrm{MeV}$ decaying exclusively into two photons (so $\text{BR}_{ee} = 0$ and $\text{BR}_{\gamma\gamma}=1$).

Running the executables, information regarding the current state of the calculation is provided and the final output is given by a 9x3 matrix comprising the final abundances after photodisintegration for the given parameter point. For our example above the corresponding output roughly looks as follows:\footnote{In case the database files have not been downloaded, the first two lines will not be present and the second step will take significantly longer.}
\begin{lstlisting}[language=bash, backgroundcolor=\color{white}]
INFO   : Extracting/Reading data files.
INFO   : Finished after 6.6s.
INFO   : Calculating non-thermal spectra and reaction rates.
INFO   : Finished after 25.1s.
INFO   : Running non-thermal nucleosynthesis.
INFO   : Finished after 25.0s.

    |    mean     |    high     |     low
----------------------------------------------
  n | 1.27835e-08 | 1.24894e-08 | 1.31176e-08
  p | 7.53096e-01 | 7.53161e-01 | 7.53036e-01
 H2 | 1.92535e-05 | 1.88105e-05 | 1.97567e-05
 H3 | 5.53065e-08 | 2.28921e-08 | 3.76285e-07
He3 | 7.63949e-06 | 3.27977e-06 | 7.82619e-06
He4 | 6.17092e-02 | 6.16965e-02 | 6.17237e-02
Li6 | 8.39044e-15 | 2.70471e-14 | 1.30028e-15
Li7 | 1.98144e-11 | 7.64333e-12 | 6.47316e-11
Be7 | 3.24951e-10 | 1.29786e-10 | 3.05213e-10
\end{lstlisting}
\vspace{5mm}
The different columns in the matrix contain the abundances $Y_N = n_N/n_b$ for the nine nuclei $N\in\{n, p, {}^2\text{H}, {}^3\text{H}, {}^3\text{He}, {}^4\text{He}, {}^6\text{Li}, {}^7\text{Li}, {}^7\text{Be}\}$ as indicated. In each column slightly different initial conditions are set,
corresponding to the SM results that are obtained by running the public code \texttt{AlterBBN}~\cite{Arbey:2011nf,Arbey:2018zfh} with the mean (first column), high (second column), and low (third column) values of the implemented nuclear reaction rates.\footnote{This can easily be changed, meaning that it is possible to use arbitrary sets of initial abundances for the calculation. We will discuss this later in section~\ref{sec:input}.}
The variation between the values of different columns can therefore be used to approximate the  theoretical errors for different abundances.

Before comparing the output of \acropolis with observationally inferred primordial abundances, it is important to note that the values returned by \acropolis do not fully take into account the three decay channels ${}^3\text{H} \rightarrow {}^3\text{He} + e^- + \bar{\nu}_e$, $n\rightarrow p + e^- + \bar{\nu}_e$, and ${}^7\text{Be} + e^- \rightarrow {}^7\text{Li} +\nu_e$, which only happen/conclude at a later cosmological time. This way, the user can in principle implement additional effects into their analysis that happen between the end of photodisintegration and the decay times. This implies that the abundances today $Y_N^* \neq Y_N$ are different than the ones returned by \acropolis and in general it is necessary to perform the following three operations
\begin{center}
\begin{tabular}{|l|l|}
	\hline
	Process & Abundance today \\
	\hline
	$n\rightarrow p + e^- + \bar{\nu}_e$ & $Y_n^* = 0\qquad Y_p^* = Y_p + Y_n$ \\
	\hline
	$ {}^3\text{H} \rightarrow {}^3\text{He} + e^- + \bar{\nu}_e$ & $Y_{^3\text{H}}^* = 0\qquad\!\! Y_{^3\text{He}}^* = Y_{^3\text{He}} + Y_{^3\text{H}}$ \\
	\hline
	${}^7\text{Be} + e^- \rightarrow {}^7\text{Li} +\nu_e$ & $Y_{^7\text{Be}}^* = 0\qquad\!\!\!\! Y_{^7\text{Li}}^* = Y_{^7\text{Be}} + Y_{^7\text{Li}}$ \\
	\hline
\end{tabular}
\end{center}
In fact, the decay of tritium might become relevant for photodisintegration reactions that happen around or beyond $\tau_{^3\text{H}} = 3.89\times 10^8\,\mathrm{s}$. However, this is at most a percent effect, since the abundance of tritium is much smaller than the one of helium-3.

\subsubsection{A first look at the source code}
While the output of the wrapper scripts is sufficient to check whether a certain parameter point is excluded or not, it is sometimes important to not only print but also to further process the final abundances. This step, however, requires some knowledge regarding the internal workings of \texttt{ACROPOLIS}, which can be obtained by taking a closer look at the actual source code. Using the file \texttt{decay} as an example, much of the important information is comprised in the following two lines of code:
\begin{lstlisting}
#! /usr/bin/env python3

[...]

from acropolis.models import DecayModel

[...]

Yf = DecayModel(mphi, tau, temp0, n0a, bree, braa).run_disintegration()

[...]
\end{lstlisting}
In this code snippet, the relevant model is first loaded from the module \texttt{acropolis.models}, which includes implementations for both example models. More precisely, the scenarios from section~\ref{sec:decay_model} (\texttt{decay}) and section~\ref{sec:annihilation_model} (\texttt{annihilation}) are implemented in the classes \texttt{acropolis.models.DecayModel} and \texttt{acropolis.models.AnnihilationModel}, respectively. After loading the model, a new instance of the respective class is initialised with the appropriate input parameters -- which here are obtained by parsing the command-line arguments --, and the calculation is initiated by calling the method \texttt{run\_disintegration()}. The latter function returns a \texttt{numpy.ndarray} of dimension $9\times n$ containing the final abundances, with $n$ being the number of different sets of initial conditions, i.e.\ $n=3$ by default. Finally, the resulting array \texttt{Yf} can either be pretty-printed (like in the wrapper scripts) or it can be used as an input for further calculations. In fact, these two lines of code are everything it takes to calculate the abundances after photodisintegration for a given model, regardless of whether the model ships with \acropolis or was implemented by the user.

\subsubsection{Running parameter scans}
\label{sec:scans}
\acropolis provides two classes for parameter scans, \texttt{acropolis.scans.BufferedScanner} and \texttt{acropolis.scans.ScanParameter}. In order to understand how these two classes work we show the following code snippet, which performs a parameter scan for the model from sec~\ref{sec:decay_model} with $m_\phi=50\,\mathrm{MeV}$, $T_0 = 10\,\mathrm{MeV}$, $\text{BR}_{ee} = 0$, and $\text{BR}_{\gamma\gamma}=1$ in the $\tau_\phi - n_\phi/n_\gamma|_{T_0}$ parameter plane.
\begin{lstlisting}
#! /usr/bin/env python3

[...]

from acropolis.models import DecayModel
from acropolis.scans  import ScanParameter, BufferedScanner

[...]

res = BufferedScanner( DecayModel,
                       mphi  = 50.,
                       tau   = ScanParameter(3, 10, 200),
                       temp0 = 10.,
                       n0a   = ScanParameter(-14, -3, 200, fast=True),
                       bree  = 0.,
                       braa  = 1.
                     ).perform_scan()

[...]
\end{lstlisting}
Here, the previously mentioned classes are first loaded from the module \texttt{acropolis.scans} together with the model that is used for the scan, in this case \texttt{DecayModel}. Then a new instance of \texttt{BufferedScanner} is initiated, which takes as a first argument the model that is used for the calculation and afterwards a set of keyword arguments with names that are identical to the ones in the constructor of the model. These keyword arguments can either be a \texttt{float} or an instance of \texttt{ScanParameter}. While the \texttt{float} parameters are kept constant, instances of \texttt{ScanParameter} are scanned over. The range for the scan is defined by the arguments of \texttt{ScanParameter(ivalue, fvalue, num, spacing)} and is constructed internally using \texttt{NumPy} functions by either calling \texttt{np.logspace(ivalue, fvalue, num)} for \texttt{spacing="log"} (default) or \texttt{np.linspace(ivalue, fvalue, num)} for \texttt{spacing="lin"}. Hence, the above code performs a scan over $\tau_\phi \in [10^3\,\mathrm{s}, 10^{10}\,\mathrm{s}]$ and $n_\phi/n_\gamma|_{T_0} \in [10^{-14}, 10^{-3}]$ with 200 data points in each direction, distributed equidistantly on a log-scale. There is one additional argument that can be passed to \texttt{ScanParameter},  \texttt{fast}, which can either be \texttt{True} or \texttt{False}. To understand the importance of this parameter, let us note that according to eq.~\eqref{eq:cascade_full}, the spectra $\kappa F_x(E)$ with some constant $\kappa$ are a solution of the cascade equation with source terms $\kappa S_x(E)$ if the spectra $F_x(E)$ are valid solutions of the cascade equation with source terms $S_x(E)$. In these two cases, the corresponding matrices in eq.~\eqref{eq:rate_matrix} are then $\kappa \mathcal{R}(T)$ and $\mathcal{R}(T)$, respectively. Hence, for each parameter that simply scales the source terms, it is not necesary to recalculate the non-thermal spectra and nuclear rates for each parameter point. Instead, we can simply rescale previously obtained solutions.
The argument \texttt{fast} is used to determine whether this procedure is used or not. In the example above with \texttt{fast=True} for \texttt{n0a} this means that for a given value of $\tau_\phi$, $\mathcal{R}(T)$ is calculated only once for $n_\phi/n_\gamma|_{T_0} = 10^{-14}$ and then simply rescaled for other values of $n_\phi/n_\gamma|_{T_0}$. This way, the initial calculation for each \texttt{tau} takes $\sim \mathcal{O}(1-10\,\mathrm{min})$, while the calculation of different points for \texttt{n0a} then merely takes $\sim \mathcal{O}(1\,\mu\mathrm{s})$.
The following tables show which parameters can be used with \texttt{fast=True} without spoiling the calculation (i.e.\ those parameters that only enter the source terms as a prefactor and do not appear elsewhere in the calculation):
\begin{center}
\begin{minipage}{0.495\textwidth}
\centering
\texttt{DecayModel}\\
\begin{tabular}{|c|c|}
	\hline
	Parameter & \texttt{fast=True} \\
	\hline
	\texttt{mphi} & No \\
	\hline
	\texttt{tau} & No \\
	\hline
	\texttt{temp0} & No \\
	\hline
	\texttt{n0a} & Yes \\
	\hline
	\texttt{bree} & No \\
	\hline
	\texttt{braa} & No \\
	\hline
\end{tabular}
\end{minipage}
\begin{minipage}{0.495\textwidth}
\centering
\texttt{AnnihilationModel}\vspace{2.5pt}\\
\begin{tabular}{|c|c|}
	\hline
	Parameter & \texttt{fast=True} \\
	\hline
	\texttt{mchi} & No \\
	\hline
	\texttt{a} & Yes (if \texttt{b=0})  \\
	\hline
	\texttt{b} & Yes (if \texttt{a=0}) \\
	\hline
	\texttt{tempkd} & No \\
	\hline
	\texttt{bree} & No \\
	\hline
	\texttt{braa} & No \\
	\hline
\end{tabular}
\end{minipage}
\end{center}
\vspace{4mm}

Finally, once the instance of \texttt{BufferedScanner} is created, the scan is initiated via a call to the method \texttt{perform\_scan()}. This function performs the scan on several cores, the number of which can be specified via the optional argument \texttt{cores}. If the latter is not specified (set to -1), only one (all available) cores are used. The given function then returns the array \texttt{res}, which contains one line for each parameter combination that was used in the scan. Each line is hence composed of the current parameter combination (two columns) and the corresponding final abundances (nine columns for each set of initial conditions).

\begin{figure}
	\centering
	\includegraphics[width=0.495\textwidth]{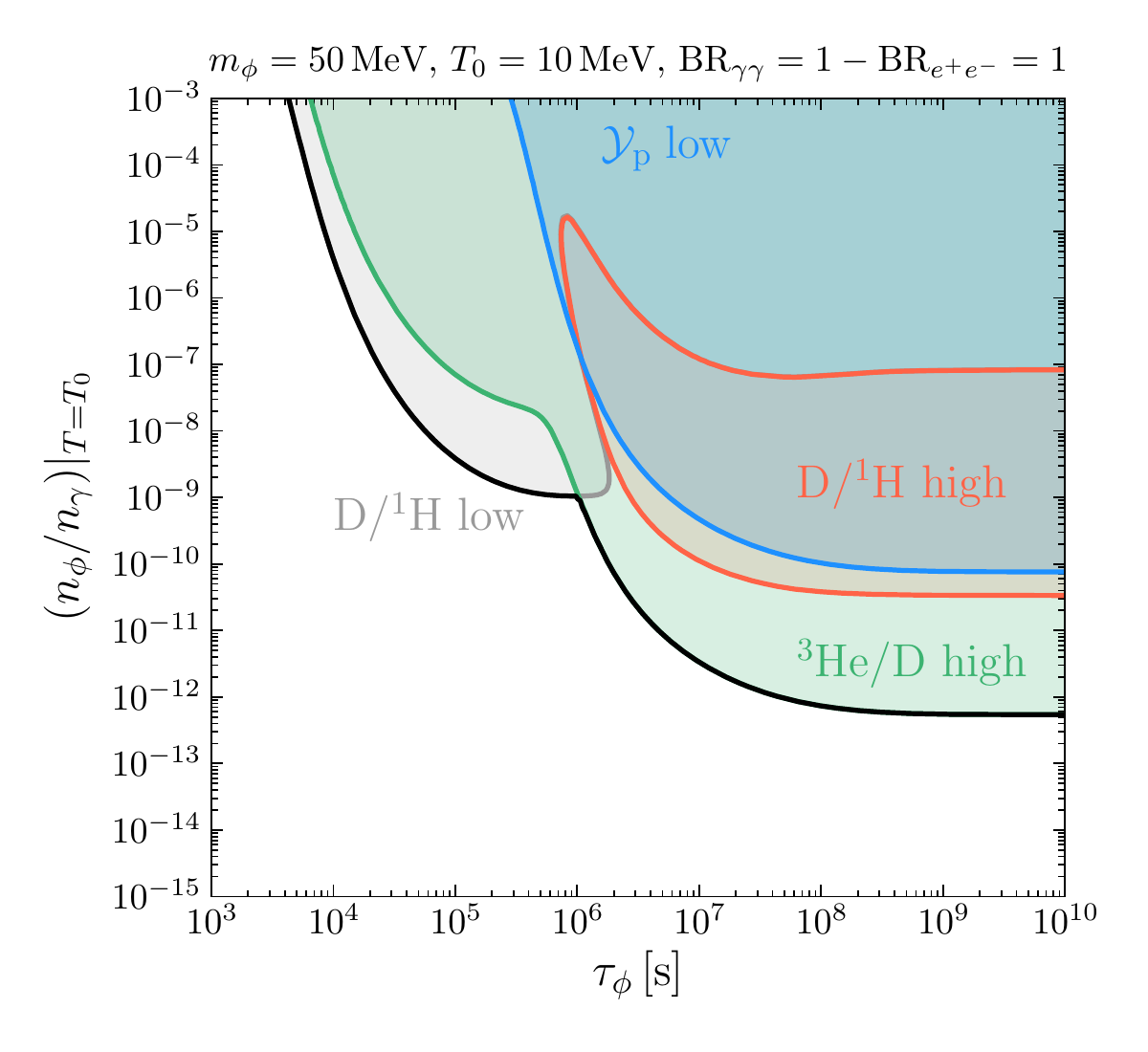}
	\includegraphics[width=0.495\textwidth]{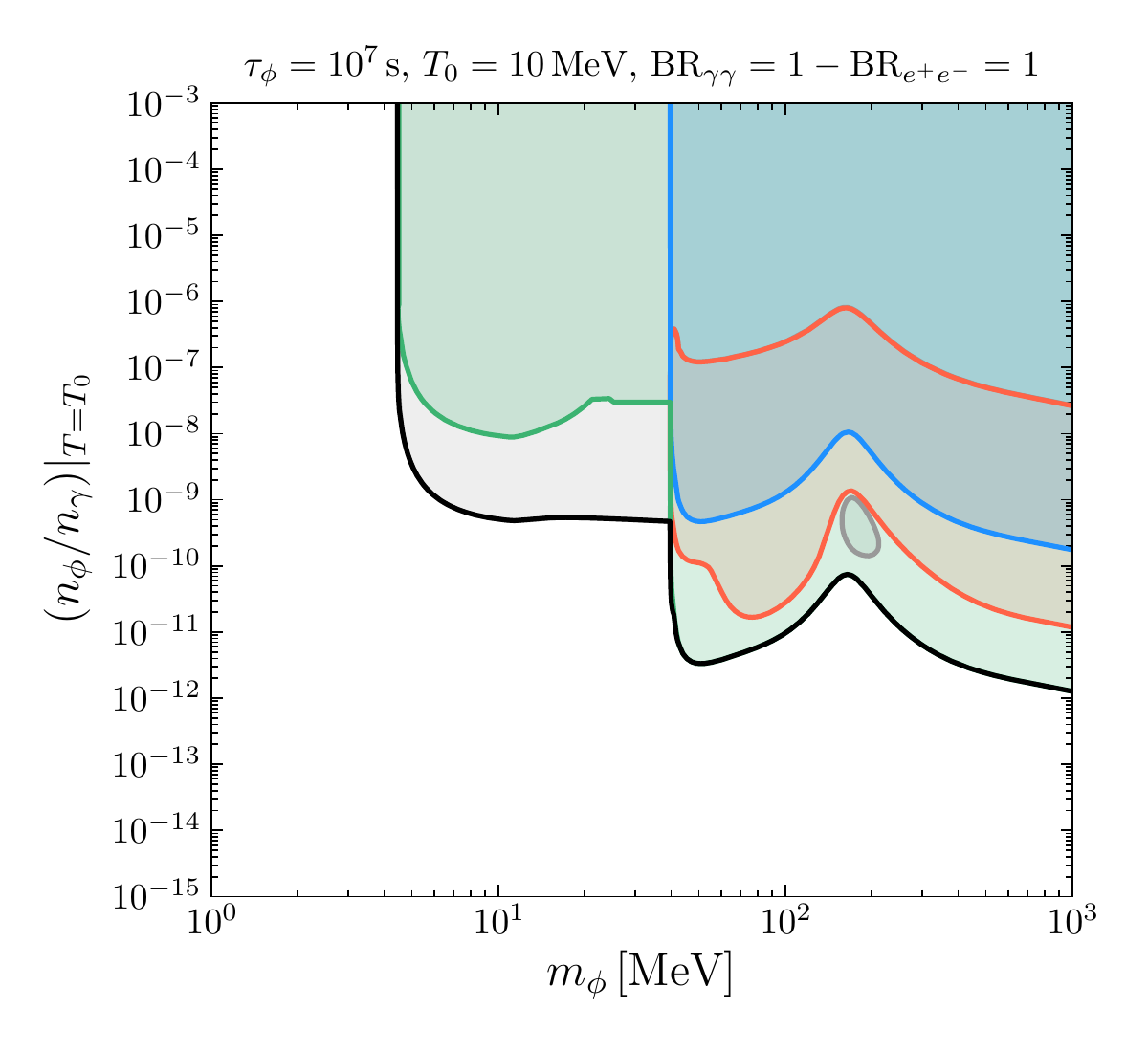}
	\caption{95\% C.L.\ constraints for decay of a decoupled MeV-scale BSM particle (implemented in \texttt{DecayModel}) into two photons ($\text{BR}_{\gamma \gamma} = 1 - \text{BR}_{e^+ e^-} = 1$) in $\tau_\phi - (n_\phi/n_\gamma)|_{T=T_0}$ plane (left) and $m_\phi - (n_\phi/n_\gamma)|_{T=T_0}$ plane (right) with $T_0 = 10 \, \mathrm{MeV}$ and $m_\phi = 50 \, \mathrm{MeV}$ (left) as well as $\tau_\phi = 10^7 \, \mathrm{s}$ (right). The limits from individual observables are shown separately: primordial deuterium abundance (orange high, grey low), helium-4 mass fraction $\mathcal{Y}_\text{p}$ (blue), and helium-3 abundance normalised by deuterium (green). The overall 95\% C.L.\ BBN limit is given by the black full line as an envelope of individual 95\% C.L.\ constraints neglecting correlations. Using $(n_\phi/n_\gamma)|_{T=T_0}$, i.e.\ \texttt{n0a}, as a \texttt{fast} parameter on a single computing node with two \texttt{AMD EPYC 7402 24-Core Processors} the scans took $\sim 40$ min (left) and $\sim 2$ h (right) for a $200 \times 200$ grid.}
	\label{fig:scan_decay}
\end{figure}

\begin{figure}
	\centering
	\includegraphics[width=0.495\textwidth]{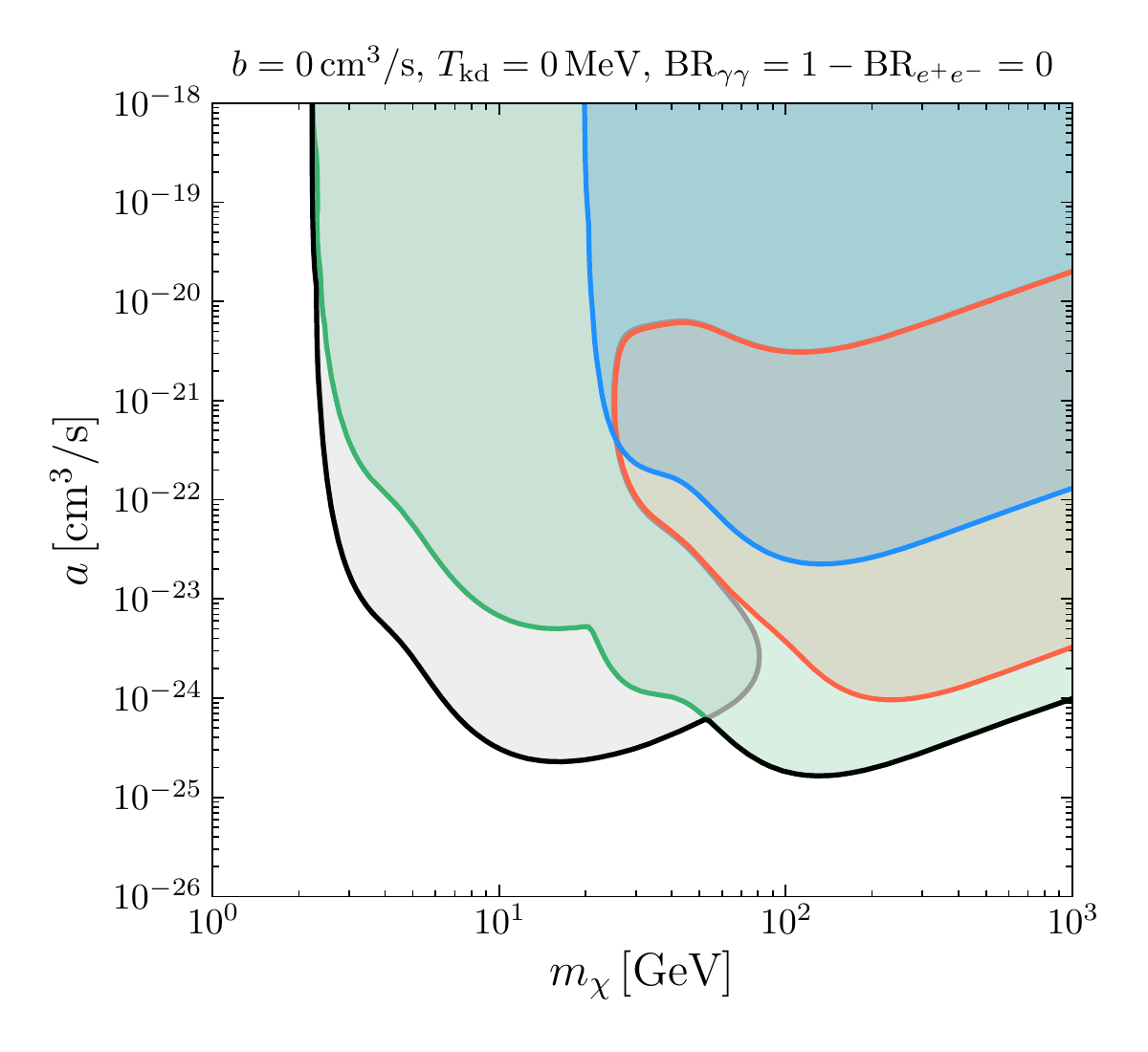}
	\includegraphics[width=0.495\textwidth]{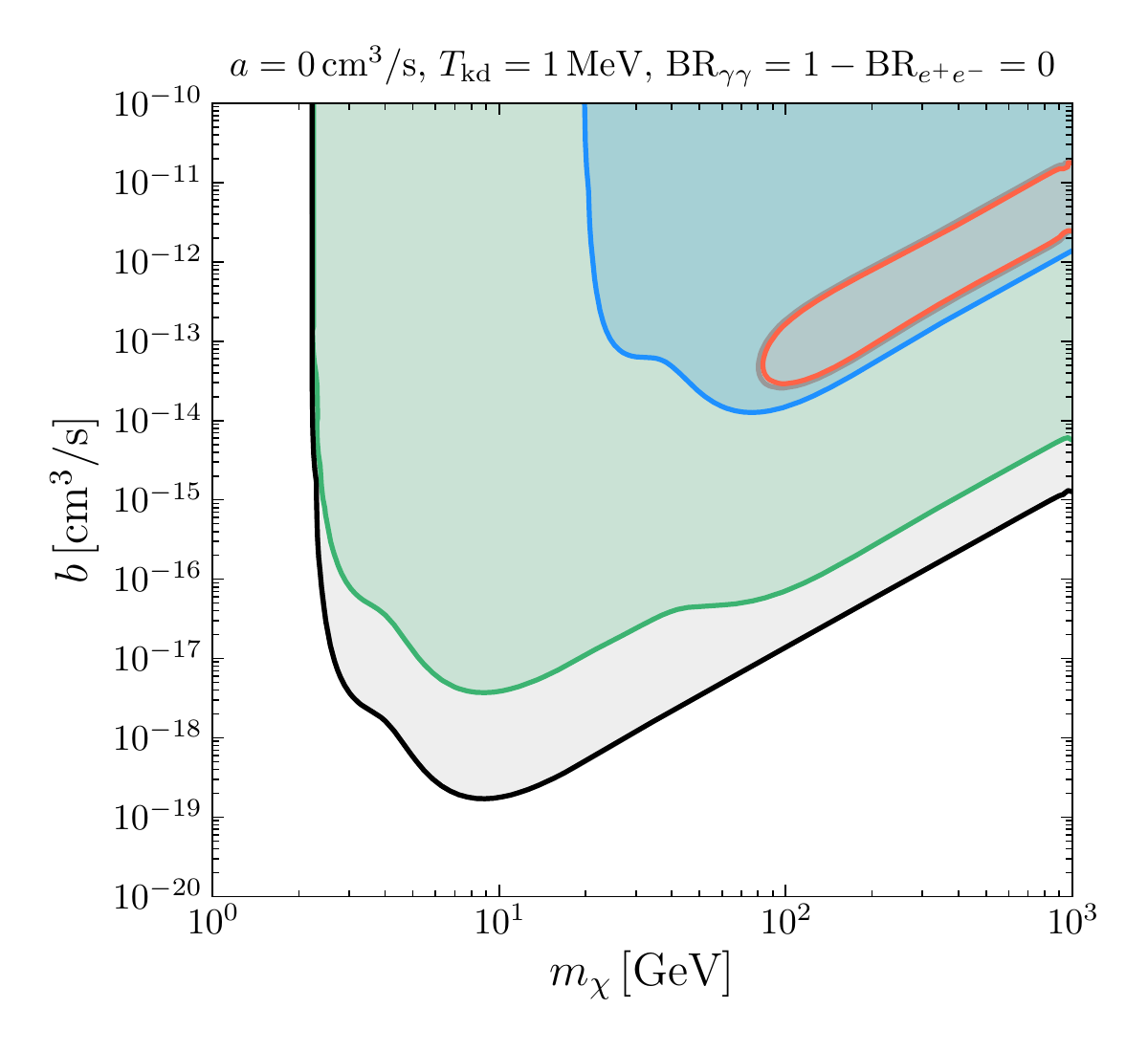}
	\caption{95\% C.L.\ constraints for residual annihilations of DM (implemented in \texttt{AnnihilationModel}) into two photons ($\text{BR}_{\gamma \gamma} = 1 - \text{BR}_{e^+ e^-} = 1$) for purely $s$-wave annihilations (left, $b = 0$) and purely $p$-wave annihilations (right, $a = 0$, $T_\mathrm{kd} = 1 \, \mathrm{MeV}$). For the explanation of the colour-coding see figure~\ref{fig:scan_decay}. Using $(n_\phi/n_\gamma)|_{T=T_0}$, i.e.\ \texttt{n0a}, as a \texttt{fast} parameter on a single computing node with two \texttt{AMD EPYC 7402 24-Core Processors} the scans took $\sim 10$ h each for a $200 \times 200$ grid.}
	\label{fig:scan_annih}
\end{figure}

To illustrate the performance of this framework, we performed several scans in both models, the results of which are shown in figures~\ref{fig:scan_decay} and \ref{fig:scan_annih}. In addition to the overall 95\% C.L.\ limit (black)
we also indicate the parts of parameter space that are excluded due to an underproduction of helium-4 (blue), over- or underproduction of deuterium (orange/grey), and overproduction of helium-3 relative to deuterium (green).\footnote{Here we adopt the latest recommendations for the observed abundances of $\mathcal{Y}_\mathrm{p} = (2.45 \pm 0.03) \times 10^{-1}$ and $\text{D}/{}^1\text{H} = (2.547 \pm 0.035) \times 10^{-5}$ from~\cite{Zyla:2020zbs}, where we took into account the uncertainty due to the baryon-to-photon ratio from Planck for $\text{D}/{}^1\text{H}$, cf.\ \cite{Aghanim:2018eyx,Depta:2020wmr}, as well as ${}^3 \text{He}/\text{D} = (8.3 \pm 1.5) \times 10^{-1}$ as an upper limit from~\cite{Geiss2003}. The nuclear rate uncertainties are taken into account as detailed in~\cite{Hufnagel:2018bjp}.}

In figure~\ref{fig:scan_decay} (left) we show the constraints for fixed $m_\phi = 50 \, \mathrm{MeV}$ in the $\tau_\phi - (n_\phi/n_\gamma)|_{T=T_0}$ plane with $T_0 = 10 \, \mathrm{MeV}$ and decays into two photons. The limits start around $\tau_\phi \sim 10^4 \, \mathrm{s}$, quickly become very stringent with increasing lifetime, and eventually flatten out excluding a number density far below the photon number density at $T_0 = 10 \, \mathrm{MeV}$. Note that due to this high sensitivity we are able to assume a standard cosmological history apart from photodisintegration as detailed in section~\ref{sec:decay_model}.
In the right panel of figure~\ref{fig:scan_decay} we show the constraints for fixed $\tau_\phi = 10^7 \, \mathrm{s}$ (right) in the $m_\phi - (n_\phi/n_\gamma)|_{T=T_0}$ plane with $T_0 = 10 \, \mathrm{MeV}$ and decays into two photons. The limits start at twice the disintegration threshold for deuterium, $m_\phi = 2 E_\mathrm{D}^\mathrm{th} \approx 4.4 \, \mathrm{MeV}$. Apart from some regions with more complex structure due to different disintegration reactions the limits become increasingly strong with larger $m_\phi$ as the energy density injected into the SM becomes larger.

The scans for figure~\ref{fig:scan_decay} took $\sim 40$ min (left) and $\sim 2$ h (right) for a $200 \times 200$ grid on an  \texttt{AMD EPYC 7402 24-Core Processors}, clearly highlighting the performance improvement due to the \texttt{fast} parameter $(n_\phi/n_\gamma)|_{T=T_0}$, i.e.\ \texttt{n0a}, making the number of points in this direction computationally inexpensive (cf.\ also appendix~\ref{sec:benchmarks_single}). The runtime is thus determined mostly by the number of points in the direction of $\tau_\phi$ or $m_\phi$ (not \texttt{fast}). Note that the longer runtime for the right panel is a result of the database files for the electromagnetic cascade reaction rates having an upper limit on the energy of $m_\phi/2 = E_0 = 100 \, \mathrm{MeV}$, which often corresponds to the most interesting region in parameter space. For masses above the pion threshold in particular, $m_\phi \gtrsim 280 \, \mathrm{MeV}$, hadrodisintegration may become relevant if $\phi$ has non-vanishing couplings to quarks, implying that $\text{BR}_{\gamma \gamma} + \text{BR}_{e^+ e^-} < 1$ in general. Also muons are kinematically available in the mass region (which are currently not implemented in \texttt{ACROPOLIS}).

In figure~\ref{fig:scan_annih} we show the constraints for residual annihilations of DM into two photons ($\text{BR}_{\gamma \gamma} = 1 - \text{BR}_{e^+ e^-} = 1$) for purely $s$-wave annihilations (left, $b = 0$) and purely $p$-wave annihilations (right, $a = 0$, $T_\mathrm{kd} = 1 \, \mathrm{MeV}$) as implemented in \texttt{AnnihilationModel}. These limits start at the disintegration threshold of deuterium, $m_\chi = E_\mathrm{D}^\mathrm{th} \approx 2.2 \, \mathrm{MeV}$, and closely resemble those presented in~\cite{Depta:2019lbe}, albeit with updated observationally inferred primordial abundances. We therefore refer to~\cite{Depta:2019lbe} for a detailed discussion. The scans took $\sim 10$ h each for a $200 \times 200$ grid on the aforementioned computing node.

\section{Implementing your own models}
\label{sec:implementing_models}

\subsection{The model framework \texttt{acropolis.models}}
While the provided example models should suffice to tackle most problems of interest, it may sometimes still happen that a scenario cannot directly be mapped to the standard implementation in \texttt{ACROPOLIS}. For such cases, \acropolis provides further tools that allow for an easy implementation of additional models.
The most important class in this context is \texttt{acropolis.models.AbstractModel}, which is an abstract base class containing most of the low-level implementation needed to run its method \texttt{run\_disintegration()}. In fact, using this class as a base, any new model can be implemented in only two steps:
\begin{enumerate}
\item[\textit{(i)}] create a new class, say \texttt{NewModel}, that uses \texttt{AbstractModel} as a base class, and
\item[\textit{(ii)}]  implement all abstract methods that are provided by \texttt{AbstractModel}, i.e.\footnote{By default, the function \texttt{AbstractModel.\_source\_positron()} simply returns the output of \texttt{AbstractModel.\_source\_electron()}, which is justified for most scenarios. However, if your specific scenario predicts different source terms for electrons and positrons, it is always possible to simply overwrite the former function.}
\begin{itemize}
	\item \texttt{AbstractModel.\_temperature\_range()}
	\item \texttt{AbstractModel.\_source\_photon()}
	\item \texttt{AbstractModel.\_source\_electron()}
	\item \texttt{AbstractModel.\_fsr\_source\_photon()}
\end{itemize}
\end{enumerate}
Consequently, each new model class must therefore feature the scaffold (compare this e.g.\ to the implementations of \texttt{DecayModel} and \texttt{AnnihilationModel})
\begin{lstlisting}
from acropolis.models import AbstractModel

class NewModel(AbstractModel):

	def _temperature_range(self):
		[...]

	def _source_photon(self, T):
		[...]

	def _source_electron(self, T):
		[...]

	def _source_photon_fsr(self, E, T):
		[...]

\end{lstlisting}
\vspace{4mm}
In the following sections, we will further discuss how to implement these four abstract methods. Apart from these methods, the constructor of \texttt{AbstractModel}, which is to be called in the constructor of \texttt{NewModel}, also needs the injection energy $E_0$ for the monochromatic part of the source term (cf.\ eq.~\eqref{eq:SXE_definition}) and an instance of the class \texttt{InputInterface} containing the necessary input data. This will be detailed in section~\ref{sec:constructor}.

\subsubsection{The functions for the source terms}
The functions \texttt{\_source\_photon} and \texttt{\_source\_electron} are associated with the respective source terms $S_\gamma^{(0)}(T)$ and $S_{e^-}^{(0)}(T)$ that enter in eq.~\eqref{eq:SXE_definition}. Both of these functions take as their only argument the temperature $T$ [in MeV] and are expected to return the corresponding source term [in $1/\mathrm{MeV}^2$]. As an example, let us take a look at \texttt{DecayModel}, which implements the source terms in eqs.~\eqref{eq:sa_decay} and \eqref{eq:se_decay} via
\begin{lstlisting}
def _source_photon(self, T):
	return self._sBRaa * 2. * self._number_density(T) * hbar/self._sTau

def _source_electron(self, T):
	return self._sBRee * self._number_density(T) * hbar/self._sTau
\end{lstlisting}
Here, \texttt{self.\_sBRee} and \texttt{self.\_sTau} are two of the input parameters of the model -- which are set in the constructor (cf.\ section~\ref{sec:constructor}) -- and \texttt{hbar} is a constant that has been imported from \texttt{acropolis.params}. We will take a closer look at the available parameters in section~\ref{sec:params}. Finally, the function \texttt{self.\_number\_density(T)} implements $n_\phi(T)$ from eq.~\eqref{eq:n_decay} and is exclusive to \texttt{DecayModel}.\footnote{While it is only necessary to implement the four abstract methods of \texttt{AbstractModel}, it is of course also possible to provide other (private) methods that are needed within the new model.}

Similarly, the function \texttt{\_source\_photon\_fsr} is associated with the final-state radiation source term $S_\gamma^\text{(FSR)}(E, T)$ entering eq.~\eqref{eq:SXE_definition}. This function takes two arguments, the energy $E$ and the temperature $T$ [both in MeV], and is expected to return the corresponding source term [in $1/\mathrm{MeV}^3$]. As an example, we show the corresponding implementation in \texttt{DecayModel} of the final-state radiation source term from eq.~\eqref{eq:safsr_decay}:
\begin{lstlisting}
def _fsr_source_photon(self, E, T):
	EX = self._sE0

	x = E/EX
	y = me2/(4.*EX**2.)

	if 1. - y < x:
		return 0.

	_sp = self._source_electron(T)

	return (_sp/EX) * (alpha/pi) * (1.+(1.-x)**2.)/x * log((1.-x)/y)
\end{lstlisting}
Here, \texttt{self.\_sE0} is the injection energy -- which is set in the constructor -- and \texttt{me2}, \texttt{alpha}, and \texttt{pi} are constants that have been imported from \texttt{acropolis.params}.

\subsubsection{The function for the temperature range}
Additionally to the source terms from the previous section, the only other function that needs to be implemented is \texttt{\_temperature\_range}, which is expected to return a two-dimensional list with the minimal and the maximal temperature spanning the range needed in eq.~\eqref{eq:YT}. Here, it is important to ensure that this range covers all temperatures for which photodisintegration is actually relevant. Coming back to \texttt{DecayModel} as an example, photodisintegration happens around the lifetime of the particle at $t\sim\tau_\phi$, and a suitable temperature range is $[10^{-3/2}T(\tau_\phi),10^{1/2}T(\tau_\phi)]$, since the bulk of photodisintegration reactions happens only for $t>\tau_\phi$. The actual implementation in \texttt{DecayModel} is given by
\begin{lstlisting}
def _temperature_range(self):
	# The number of degrees-of-freedom to span
	mag = 2.
	# Calculate the approximate decay temperature
	Td = self._sII.temperature( self._sTau )
	# Calculate Tmin and Tmax from Td
	Td_ofm = log10(Td)
	# Here we choose -1.5 (+0.5) orders of magnitude
	# below (above) the approx. decay temperature,
	# since the main part happens after t = \tau
	Tmin = 10.**(Td_ofm - 3.*mag/4.)
	Tmax = 10.**(Td_ofm + 1.*mag/4.)

	return (Tmin, Tmax)
\end{lstlisting}
Apart from the previously discussed model parameter \texttt{self.\_sTau}, this implementation only involves the variable \texttt{self.\_sII}, which is set in the constructor (cf.\ section~\ref{sec:constructor}) and constitutes an instance of the class \texttt{acropolis.input.InputInterface}. This class is the second most important class next to \texttt{acropolis.models.AbstractModels} as it provides an interface for all the files that are used as an input for the calculation. For example, this class wraps functions like $T(t)$ (i.e.\ \texttt{InputInterface.temperature} from above), which are relevant for eqs.\ like \eqref{eq:y_pdi} and therefore need to be provided as an input. In section~\ref{sec:input} we will discuss this class in more detail and also go over the different inputs that are required for a successful calculation.

\subsubsection{The model constructor}
\label{sec:constructor}

When calling the constructor of a new model, it crucial to also invoke the constructor of the abstract base class \texttt{AbstractModel}. The latter one takes two arguments, the first one being the injection energy $E_0$ for the monochromatic part of the source term (cf.\ eq.~\eqref{eq:SXE_definition}) and the second one being an instance of the previously mentioned class \texttt{InputInterface}. When considering \texttt{DecayModel} as an example, which features an injection energy $E_0 = m_\phi/2$, this leads to a constructor of the following form
\begin{lstlisting}
def __init__(self, mphi, tau, temp0, n0a, bree, braa):
	# Initialize the Input_Interface
	self._sII   = InputInterface("data/sm.tar.gz")

	[...]

	# The injection energy
	self._sE0   = mphi/2.

	[...]

	# Call the super constructor
	super(DecayModel, self).__init__(self._sE0, self._sII)
\end{lstlisting}
Here, an instance of \texttt{InputInterface} is constructed from the data that is stored in the file \texttt{data/sm.tar.gz}. To fully understand the code, we therefore have to discuss the importance of this class, which we do in the next section.

\subsection{The input framework \texttt{acropolis.input}}
\label{sec:input}
The calculation of the abundances after photodisintegration cannot proceed without certain (model-dependent) inputs, including the baryon-to-photon ratio, the initial abundances after BBN, and the dynamics of the background plasma as encoded in functions such as $T(t)$, $T_\nu(T)$, and $H(T)$. All of these inputs are collectively wrapped by and accessible via the previously mentioned class \texttt{acropolis.input.InputInterface}. The constructor of this class only takes a single argument, which is expected to be the name of a \texttt{.tar.gz} file containing the following three files
\begin{lstlisting}[language=bash,backgroundcolor=\color{white}]
input.tar.gz
|
|___ param_file.dat
|___ cosmo_file.dat
|___ abundance_file.dat
\end{lstlisting}
One such file that ships with \acropolis is \texttt{data/sm.tar.gz}, which assumes a standard cosmological history without an appreciable impact of BSM physics apart from photodisintegration.
As mentioned above this is a very good approximation for parameter regions which are close to the resulting limit, as photodisintegration strongly constrains even very small abundances, implying that changes to the SM values are small. Correspondingly, this file is used for both example models, as in these cases the abundance of the participating dark-sector particles gives a negligible contribution.
However, this might not always be the case, potentially necessitating the construction of model-specific input files.
We go over these different \texttt{*.dat} files and their content below and discuss how to access them from an instance of \texttt{InputInterface}. This is crucial since this class might encode vital information that is needed to implement the different source terms and the temperature range.

\subsubsection{The file \textit{param\_file.dat}}
This file contains all input parameters that cannot be provided in the model constructor, e.g.\ if they are closely tied to the evolution of the background plasma. Here, the minimal requirement is to provide a value for the baryon-to-photon ratio,\footnote{This parameter also enters the initial abundances and therefore is closely tied to other inputs.} but more parameters can be incorporated via additional lines of the form \texttt{key=value}. For example, the file \texttt{data/sm.tar.gz:param\_file.dat} only contains one line,
\begin{lstlisting}[backgroundcolor=\color{white}]
eta=6.137e-10
\end{lstlisting}
However, independent of the number of lines in this file, after constructing an instance of \texttt{InputInterface}, all included parameters can be accessed by simply calling the method \texttt{InputInterface.parameter(key)} with the corresponding \texttt{key}. For example in the case of \texttt{sm.tar.gz}, calling \texttt{parameter("eta")} would return \texttt{6.137e-10}.

\subsubsection{The file \textit{cosmo\_file.dat}}
This file encodes the cosmological evolution of the background plasma. It must contain at least five columns (separated by spaces) including discrete grids for the following quantities:
\begin{itemize}
\item time $t$ [in s]
\item temperature $T$ [in MeV]
\item time-temperature relation $\text{d}T/\text{d}t$ [in MeV$^2$]
\item neutrino temperature $T_\nu$ [in MeV]
\item Hubble rate $H$ [in MeV]
\end{itemize}
The grid needs to be equidistantly spaced on a log-scale of time $t$. Given an instance of \texttt{InputInterface}, the interpolated data of these mandatory entries is accessible via the predefined methods
\begin{center}
\begin{tabular}{rl}
	$T(t)$ & \texttt{InputInterface.temperature(t)} \\[0.2cm]
	$t(T)$ & \texttt{InputInterface.time(T)} \\[0.2cm]
	$\frac{\text{d}T}{\text{d}t}(T)$ & \texttt{InputInterface.dTdt(T)} \\[0.2cm]
	$T_\nu(T)$ & \texttt{InputInterface.neutrino\_temperature(T)} \\[0.2cm]
	$H(T)$ & \texttt{InputInterface.hubble\_rate(T)} \\[0.2cm]
	$R(T)$ & \texttt{InputInterface.scale\_factor(T)} \\[0.2cm]
\end{tabular}
\end{center}

\noindent Besides these mandatory columns, it is also possible to add an arbitrary number of additional ones, e.g.\ containing quantities that are needed to implement the source terms or the temperature range. The interpolated data of the columns can then be accessed by calling the more generic method \texttt{InputInterface.cosmo\_column(yc, val, xc=1)}. This function can be used to generically calculate $y(x)$ at $x=$\texttt{val} via interpolation of any two columns in \textit{cosmo\_file.dat}, where $y$ is determined from the entries in column \texttt{yc} and $x$ is determined from the entries in column \texttt{xc}. This function therefore does not only allow to evaluate user-defined columns but also different correlations between the mandatory quantities, like e.g.\ $t(H)$. Possible examples include

\begin{center}
\begin{tabular}{|c|c|c|}
	\hline
	\texttt{yc} & \texttt{xc} & $y(x)$ \\
	\hline
	1 & 0 & $T(t)$ \\
	\hline
	4 & 3 & $H(T_\nu)$ \\
	\hline
	0 & 4 & $t(H)$ \\
	\hline
	5 & 1 & $C_5(T)$ \\
	\hline
\end{tabular}
\end{center}

\noindent In the last example, column \texttt{yc=5} is assumed to exist (i.e.\ that it was added by the user), in which case $C_5$ denotes the cosmological quantity that is tabulated in this column.

\subsubsection{The file \textit{abundance\_file.dat}}
This file contains the initial abundances (before photodisintegration, but after nucleosynthesis) that are used for the calculation. This file must contain at least one column with the abundances for $n$, $p$, ${}^2\text{H}$, ${}^3\text{H}$, ${}^3\text{He}$, ${}^4\text{He}$, ${}^6\text{Li}$, ${}^7\text{Li}$, and ${}^7\text{Be}$. Additional columns are also allowed and -- as previously mentioned -- the code calculates the resulting abundances after photodisintegration for each set of initial values (i.e.\ for each column). After constructing an instance of \texttt{InputInterface}, the different initial abundances can be collectively accessed by calling the method \texttt{InputInterface.bbn\_abundances()}, which returns a $9\times n$ array with $n$ being the number of columns. Alternatively, the first column can also be separately obtained by calling \texttt{InputInterface.bbn\_abundance\_0()} instead.

\subsection{The parameters in \texttt{acropolis.params}}
\label{sec:params}
All of the constants that are used within \acropolis can be found in \texttt{acropolis.params}. This files not only contains mathematical (\texttt{zeta3},...) and physical (\texttt{alpha, me, hbar},...) constants, but also some algorithm-specific parameters such as the number of points per decade used to construct the energy (\texttt{NE\_pd}) and temperature (\texttt{NT\_pd}) grids. All of these parameters are well documented in the file itself.
There are three parameters at the beginning of this file that can be set to \texttt{True} or \texttt{False}:
\begin{itemize}
	\item \texttt{verbose} (default: \texttt{True}) -- if this parameter is set to \texttt{True}, the code prints all message types to the screen, including \texttt{INFO}, \texttt{WARNING}, and \texttt{ERROR}. If this parameter is set to \texttt{False} instead, \texttt{INFO} messages are not printed.
	\item \texttt{debug} (default: \texttt{False}) -- if this parameter is set to \texttt{True}, additional debug info is printed, including additional information on the position at which certain warnings or errors appear.
	\item \texttt{usedb} (default: \texttt{True}) -- if this parameter is set to \texttt{True}, the optional database files are used to speed up the calculation at the cost of a higher RAM usage. If this parameter is set to \texttt{False}, all reaction rates are calculated from scratch, regardless of whether the database files have been downloaded or not.
\end{itemize}

\begin{figure}
	\includegraphics[width=0.495\textwidth]{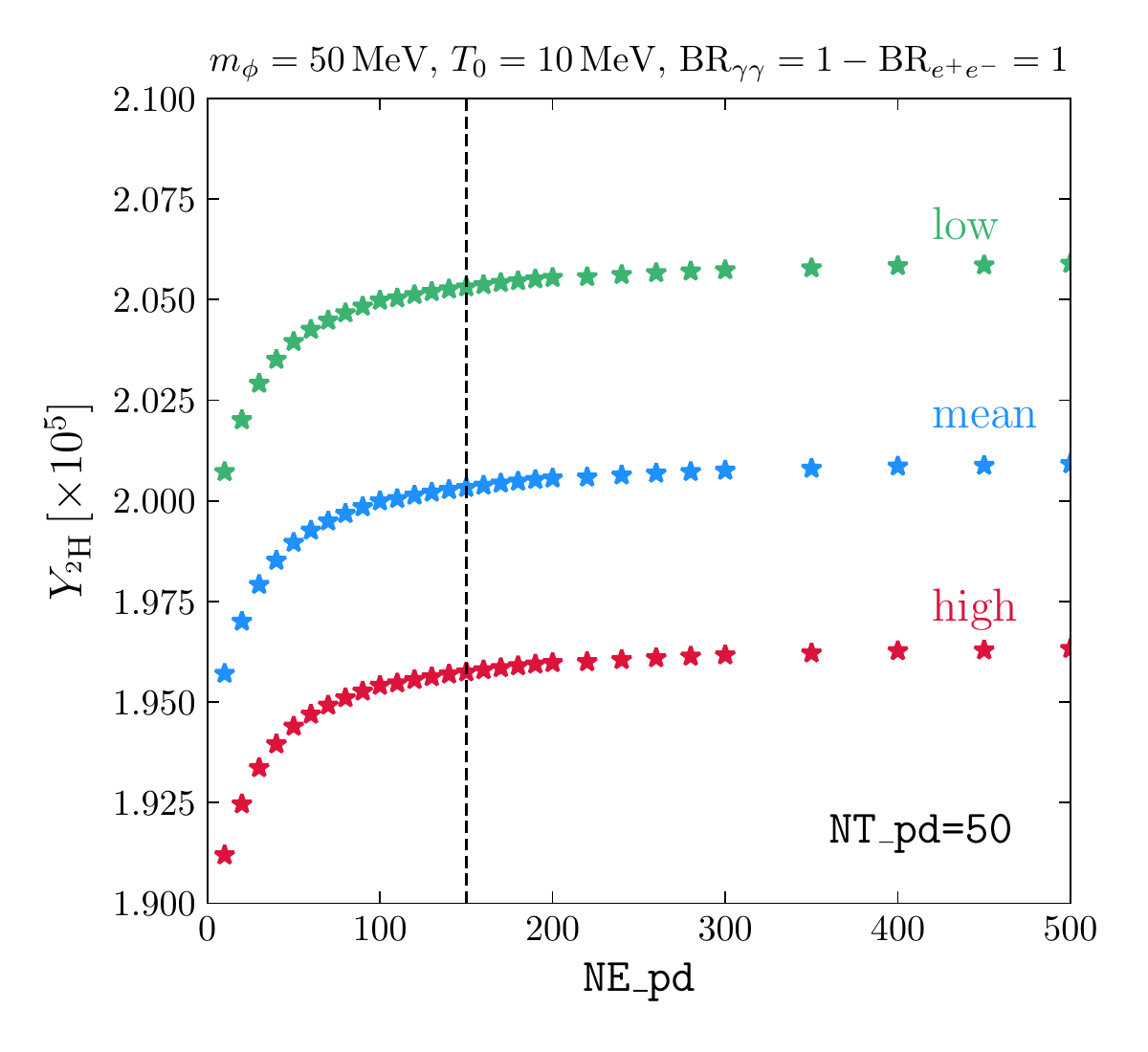}
	\includegraphics[width=0.495\textwidth]{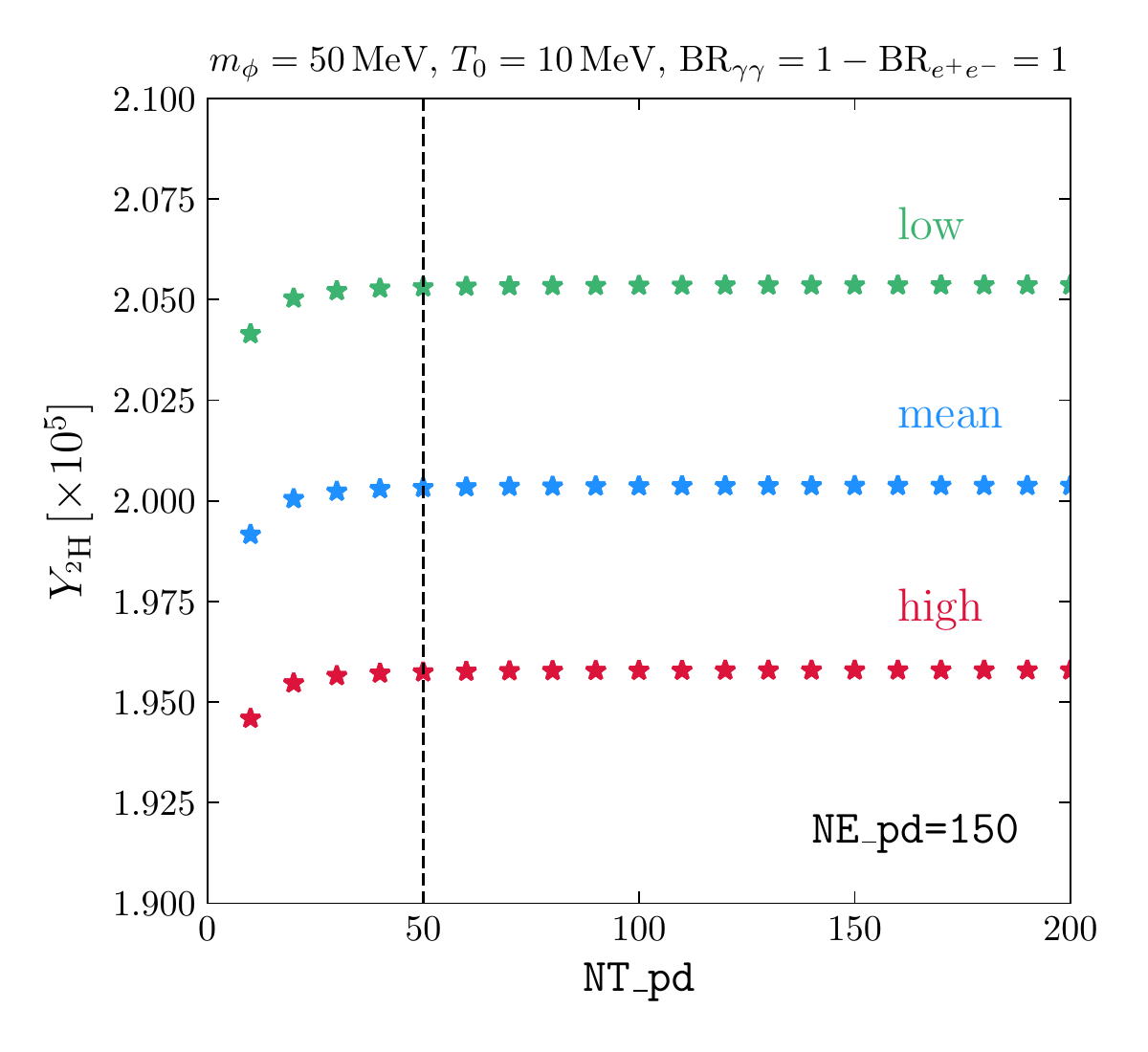}
	\caption{Convergence of the abundance of deuterium as a function of the grid points \texttt{NE\_pd} and \texttt{NT\_pd}. The dashed line indicates the default value in \texttt{ACROPOLIS}.}
	\label{fig:convergence}
\end{figure}

For all other parameters, it is not advised to perform any manual changes, since these were selected in a way to ensure a great compromise between runtime and accuracy around the exclusion region.\footnote{Far away from the exclusion line, the results might not be perfectly accurate. If, for some reason, you also require precise results in this region, try increasing the values of \texttt{NE\_pd} and \texttt{NT\_pd}.} We demonstrate this in figure~\ref{fig:convergence}, where we show the final abundance of deuterium for different values of \texttt{NE\_pd} and \texttt{NT\_pd}.
We find that the abundances indeed converge for \texttt{NE\_pd},\texttt{NT\_pd}$\rightarrow \infty$, while the default values for \texttt{NE\_pd} and \texttt{NT\_pd} (indicated by the dashed lines) lead to a result that deviates only at the 0.1\% level
-- much smaller than the difference that is caused by the reaction rate uncertainties (different colours).

\section{Conclusions}

In this work, we present \texttt{ACROPOLIS}, \texttt{A} generi\texttt{C} f\texttt{R}amework f\texttt{O}r \texttt{P}hotodisintegration \texttt{O}f \texttt{LI}ght element\texttt{S}, the first public code for calculating the effects of photodisintegration of light elements in the early universe. \acropolis performs this by first computing the non-thermal part of the photon spectrum arising due to late-time high-energetic injections into the SM plasma and then calculating its effect on the primordial light element abundances. We provide and discuss two example programs covering a plethora of interesting applications via {\it (i)} the decay of a decoupled MeV-scale BSM particle and {\it (ii)} residual annihilations of DM. Furthermore, we detail how additional models can easily be implemented in the modular structure of \texttt{ACROPOLIS}.

\acknowledgments

This work is supported by the ERC Starting Grant `NewAve' (638528), the
Deutsche Forschungsgemeinschaft under Germany's Excellence Strategy -- EXC 2121 `Quantum Universe' -- 390833306,
and by the F.R.S.\ FNRS under the Excellence of Science (EoS) project No.\ 30820817 be.h `The H boson gateway to physics beyond the Standard Model'.

\appendix

\section{Rates for the cascade processes}
\label{app:rates_cascade}
\newcommand{\be}{\bar{\epsilon}}
\newcommand{\Ep}{E' + \be - E}

In this appendix, we collect for completeness all relevant total and differential interaction rates $\Gamma_\X(E)$ and $K_{\X' \to \X}(E,E')$ for the cascade processes of high-energetic photons, electrons, and positrons on the background photons, electrons, and nuclei (see eqs.~\eqref{eq:cascade_f} and~\eqref{eq:cascade_full}). Large parts are directly taken from~\cite{Hufnagel:2018bjp}.

\subsection*{Target densities}
The thermal photon spectrum differential in energy $\fpdi_\gamma(\be)$ is given by
\begin{align}
\fpdi_\gamma(\be) = \frac{\be^2}{\pi^2} \times \frac{1}{\exp(\be/T) - 1}\eqsp,
\end{align}
while the total baryon number density can be calculated from the baryon-to-photon ratio $\eta$ and the number density of photons $n_\gamma(T)$,
\begin{align}
n_b(T) = \eta \times n_\gamma(T) = \eta \times \frac{2\zeta(3)}{\pi^2} T^3 \eqsp.
\end{align}
Via charge neutrality we obtain for the number density of background electrons
\begin{align}
\eq{n}_e(T) = \sum_N Z_Nn_N \simeq \big[Y_{p}(T) + 2 Y_{{}^4\text{He}}(T)\big]\times n_b(T),\quad Y_N(T) = \frac{n_N(T)}{n_b(T)} \eqsp.
\end{align}
At the times relevant to photodisintegration $(t \gtrsim 10^4\,\text{s})$, BBN has already terminated and the nuclear abundances $Y_N(T)$ are approximately constant. Hence, in the following, we neglect the temperature dependence of $Y_N(T)$, and fix them to their values directly after BBN. Note that a change due to photodisintegration is only relevant when the corresponding scenario is excluded anyhow.

\subsection*{Final-state radiation: $\boldsymbol{\text{DS} \rightarrow e^+ e^- \gamma}$}

Following \cite{Forestell:2018txr}, the source term for final-state radiation can directly be calculated from the source term of electrons or positrons via the expression \cite{Mardon:2009rc,Birkedal:2005ep}
\begin{equation}
S_\gamma^\text{(FSR)}(E) = \frac{S_{e^\pm}^{(0)}}{E_0} \times \frac{\alpha}{\pi} \frac{1+(1-x)^2}{x}\ln\left( \frac{4E_0^2(1-x)}{m_e^2} \right) \times \Theta\left( 1 - \frac{m_e^2}{4E_0^2} - x \right)
\label{eq:SFSR}
\end{equation}
with $x=E/E_0$.

\subsection*{Double photon pair creation: $\boldsymbol{\gamma \gamma_\text{th} \rightarrow e^+ e^-}$}
The rate for double photon pair creation is given by~\cite{Kawasaki:1994sc}\footnote{Correcting a typo in eq.~(27) of~\cite{Kawasaki:1994sc}.}
\begin{align}
\Gamma_\gamma^\text{(DP)}(E) = \frac{1}{8E^2}\times \int_{m_e^2/E}^{\infty} \text{d}\be\; \frac{\fpdi_\gamma(\be)}{\be^2} \times \int_{4m_e^2}^{4E\be} \text{d}s\;s\cdot\sigma_\text{DP}\left(\beta = \sqrt{1 - 4m_e^2/s}\right)
\end{align}
with the total cross-section
\begin{align}
\sigma_\text{DP}(\beta) = \frac{\pi\alpha^2}{2 m_e^2} \times (1-\beta^2)\left[ (3-\beta^4)\ln\left( \frac{1+\beta}{1-\beta} \right) - 2\beta \left( 2 - \beta^2 \right) \right]\eqsp.
\end{align}
This process is only relevant above the threshold of production of electron-positron pairs $E \gtrsim m_e^2/(22T)$, allowing us to set $\Gamma_\gamma^\text{(DP)}(E) = 0$ for $E < m_e^2/(22T)$.

The differential rate for double photon pair creation entering the calculation of the electron and positron spectrum\footnote{Here, the notation $\gamma \rightarrow e^\pm$ in the index of $K_{\X' \rightarrow \X}$ indicates that the corresponding expression is valid for $\X' \rightarrow \X \in \{\gamma \rightarrow e^+, \gamma \rightarrow e^-\}$ and consequently enters eq.~\eqref{eq:cascade_full} twice.} was originally calculated in~\cite{1983Afz....19..323A} and is given by\footnote{Correcting a typo in eq.~(28) of~\cite{Kawasaki:1994sc}.}
\begin{align}
K_{\gamma \rightarrow e^\pm}^\text{(DP)}(E, E') = \frac{\pi\alpha^2 m_e^2}{4} \times \frac{1}{E'^3}\int_{m_e^2/E'}^{\infty} \text{d}\be\;\frac{\fpdi_\gamma(\be)}{\be^2}\;G(E, E', \be)\eqsp,
\end{align}
with
\begin{align}
G(E, E', \be) & = \frac{4(E' + \be)^2}{E(\Ep)}\ln\left( \frac{4\be E (\Ep)}{m_e^2(E' + \be)} \right) \nonumber \\ \nonumber \\
& + \left( \frac{m_e^2}{\be (E' + \be)} - 1 \right)\frac{(E'+\be)^4}{E^2(\Ep)^2} \nonumber \\ \nonumber \\
& + \frac{2\left[ 2\be (E' + \be) -m_e^2 \right](E'+\be)^2}{m_e^2 E (\Ep)} - 8\frac{\be (E'+\be)}{m_e^2}
\end{align}
for $m_e < E_\text{lim}^- < E < E_\text{lim}^+$,
\begin{align}
2E_\text{lim}^\pm = E'+\be \pm (E'-\be)\sqrt{1- \frac{m_e^2}{E'\be}}\eqsp,
\end{align}
and $G(E, E', \be) = 0$ otherwise. As explained above, we further set $K_{\gamma \rightarrow e^\pm}^\text{(DP)}(E, E') = 0$ for $E' < m_e^2/(22T)$.

\subsection*{Photon-photon scattering: $\boldsymbol{\gamma \gamma_\text{th} \rightarrow \gamma \gamma}$}
The total and differential interaction rates for photon-photon scattering have been originally calculated in~\cite{Svensson:1990pfo}, and are given by\footnote{Correcting a typo in eq.~(31) of~\cite{Kawasaki:1994sc} and in eq.~(5) of~\cite{Poulin:2015opa}.}
\begin{align}
\Gamma_{\gamma}^{\text{(PP)}}(E) = \frac{1946}{50625\pi}\times \frac{8\pi^4}{63}\times \alpha^4 m_e \times \left( \frac{E}{m_e} \right)^3 \left( \frac{T}{m_e} \right)^6 \eqsp,
\end{align}
and
\begin{align}
K_{\gamma \rightarrow \gamma}^\text{(PP)}(E, E') = \frac{1112}{10125\pi}\times\frac{\alpha^4}{m_e^8}\times \frac{8\pi^4T^6}{63} \times E'^2\left[ 1 - \frac{E}{E'} + \left( \frac{E}{E'} \right)^2 \right]^2 \eqsp.
\end{align}
In principle, these expressions are only valid for $E \lesssim m_e^2/T$~\cite{Kawasaki:1994sc}. However, for energies larger than this, photon-photon scattering is in any case negligible compared to double photon pair creation, making it unnecessary to impose this additional constraint.

\subsection*{Bethe-Heitler pair creation: $\boldsymbol{\gamma N \rightarrow N e^+ e^-}$}
The total rate for Bethe-Heitler pair creation at energies $E \geq 4m_e$ and up to order $m_e^2/E^2$ can be written as~\cite{Maximon:1968, Kawasaki:1994sc}\footnote{We checked that higher order terms do not change the final results.}
\begin{eqnarray}
&&\Gamma_{\gamma}^{\text{(BH)}}(E) \simeq \frac{\alpha^3}{m_e^2} \times \Bigg( \sum_N Z_N^2 n_N(T) \Bigg) \times \Bigg( \left[ \frac{28}{9}\ln(2k) -\frac{218}{27} \right] \nonumber \\
&& \qquad+ \left(\frac{2}{k}\right)^2 \left[ \frac{2}{3}\ln(2k)^3 - \ln(2k)^2 + \left( 6 - \frac{\pi^2}{3} \right)\ln(2k) + 2\zeta(3) + \frac{\pi^2}{6} - \frac{7}{2}\right]\Bigg)\Bigg|_{k=E/m_e} \eqsp.
\end{eqnarray}
Here, we only take into account scattering off $^1\mathrm{H}$ and $^4\mathrm{He}$, which implies
\begin{align}
\sum_{N} Z_N^2 n_N(T) \simeq \sum_{N\,\in\,\{p, {}^4\text{He} \}} Z_N^2 n_N(T) = \big[Y_{p}(T) + 4 Y_{{}^4\text{He}}(T)\big] \times n_b(T)\eqsp,
\end{align}
since the abundances of all other nuclei are strongly suppressed. Furthermore, for energies in the range $2m_e < E \leq 4\e\mathrm{MeV}$, the interaction rate is essentially constant~\cite{Jedamzik:2006xz}, $\Gamma_{\gamma}^{\text{(BH)}}(E) \simeq \Gamma_{\gamma}^{\text{(BH)}}(E = 4\e\mathrm{MeV})$.

The differential rate for Bethe-Heitler pair creation is given by~\cite{berestetskii1982quantum, Kawasaki:1994sc}
\begin{align}
K_{\gamma \rightarrow e^\pm}^\text{(BH)}(E, E') = \Bigg( \sum_N Z_N^2 n_N(T) \Bigg) \times \frac{\text{d}\sigma_\text{BH}(E, E')}{\text{d}E} \times \Theta(E' - E-m_e) \eqsp,
\label{kernel_BH_electron}
\end{align}
with the differential cross-section
\begin{align}
\frac{\text{d}\sigma_\text{BH}(E, E')}{\text{d}E} = & \frac{\alpha^3}{m_e^2} \times \left( \frac{p_+ p_-}{E'^3} \right) \times \Bigg[ -\frac43 - 2E_+E_- \frac{p_+^2 + p_-^2}{p_+^2 p_-^2} \nonumber \\ \nonumber \\
& + m_e^2 \left( l_-\frac{E_+}{p_-^3} + l_+\frac{E_-}{p_+^3} - \frac{l_+ l_-}{p_+ p_-} \right) \nonumber \\ \nonumber \\
& + L\left( -\frac{8E_+E_-}{3p_+p_-} + \frac{E'^2}{p_+^3 p_-^3} \left( E_+^2E_-^2 + p_+^2 p_-^2 - m_e^2E_+ E_- \right) \right) \nonumber \\ \nonumber \\
& - L\frac{m_e^2E'}{2p_+ p_-}\left( l_+\frac{E_+ E_- - p_+^2}{p_+^3} + l_-\frac{E_- E_+ - p_-^2}{p_-^3} \right) \Bigg] \eqsp,
\end{align}
where we have defined
\begin{align}
E_- \ldefine E\eqsp,\qquad E_+ \ldefine E' - E&\eqsp,\qquad p_\pm \ldefine \sqrt{E_\pm^2 - m_e^2} \\ \nonumber \\
L \ldefine \ln\left( \frac{E_+ E_- + p_+ p_- + m_e^2}{E_+ E_- - p_+ p_- + m_e^2} \right)&\eqsp, \qquad l_\pm \ldefine \ln\left( \frac{E_\pm + p_\pm}{E_\pm - p_\pm} \right) \eqsp. \\ \nonumber
\end{align}
The $\Theta$-function appearing in eq.~\eqref{kernel_BH_electron} ensures that we fulfill energy conservation in the integration of $E'$ over the range $[E, \infty]$ in eq.~\eqref{eq:cascade_full}.

\subsection*{Compton scattering: $\boldsymbol{\gamma e^-_\text{th} \rightarrow \gamma e^-}$}
The total rate for Compton scattering can be found in~\cite{Kawasaki:1994sc, Poulin:2015opa} and is given by
\begin{align}
\Gamma_\gamma^\text{(CS)}(E) = \frac{2\pi\alpha^2}{m_e^2}\times\eq{n}_e(T)\times\frac{1}{x}\left[ \left(1 - \frac{4}{x} - \frac{8}{x^2}\right)\ln(1+x) + \frac12 + \frac8x - \frac{1}{2(1+x)^2}\right]\Bigg|_{x=2E/m_e} \eqsp.
\end{align}
Furthermore, the differential rate for the energy of the scattered photon reads~\cite{Kawasaki:1994sc, Poulin:2015opa}\footnote{Correcting a typo in eq.~(10) of~\cite{Poulin:2015opa}.}
\begin{align}
K_{\gamma \rightarrow \gamma}^\text{(CS)}(E, E') &= \Theta(E-E'/(1+2E'/m_e)) \times  \frac{\pi\alpha^2}{m_e}\times \eq{n}_e(T)\times \nonumber \\
&\frac{1}{E'^2}\left[ \frac{E'}{E} + \frac{E}{E'} + \left( \frac{m_e}{E} - \frac{m_e}{E'} \right)^2 - 2m_e\left( \frac{1}{E} - \frac{1}{E'} \right) \right]
\label{kernel_CS_gamma}
\end{align}
with the $\Theta$-function corresponding to a vanishing rate above the Compton edge.

Following~\cite{Kawasaki:1994sc}, the differential rate relevant for the spectrum of electrons can be deduced from eq.~\eqref{kernel_CS_gamma},
\begin{align}
K_{\gamma \rightarrow e^-}^\text{(CS)}(E, E') = K_{\gamma \rightarrow \gamma}^\text{(CS)}(E' + m_e - E, E')\eqsp.
\end{align}

\subsection*{Inverse Compton scattering: $\boldsymbol{e^\pm \gamma_\text{th} \rightarrow e^\pm \gamma}$}
The differential rate for production of photons from inverse Compton scattering was originally calculated in~\cite{Jones:1968zza} and can be written as
\begin{align}
K_{e^\pm \rightarrow \gamma}^\text{(IC)}(E, E') = 2\pi\alpha^2\times \frac{1}{E'^2} \int_{0}^{\infty} \text{d}\be\;\frac{\fpdi_\gamma(\be)}{\be}\;F(E, E', \be) \times \Theta(E' - E - m_e)\eqsp.
\label{kernel_IC_gamma}
\end{align}
For $\be \leq E \leq 4\be E'^2/(m_e^2 + 4\be E')$, the function $F(E, E', \be)$ is given by\footnote{Correcting a typo in eq.~(49) of~\cite{Kawasaki:1994sc}.}
\begin{align}
F(E, E', \be) = 2q\ln(q) + (1+2q)(1-q)+\frac{\Gamma_\epsilon^2 q^2}{2+2\Gamma_\epsilon q}(1-q) \eqsp,
\end{align}
with
\begin{align}
\Gamma_\epsilon = \frac{4\be E'}{m_e^2}\eqsp, \qquad q = \frac{E}{\Gamma_\epsilon(E' - E)}\eqsp,
\end{align}
and $F(E, E', \be) = 0$ otherwise.\footnote{According to~\cite{Jones:1968zza}, the function $F(E, E', \be)$ takes a different form for $E < \be$. However, this part of parameter space is practically irrelevant for our considerations.} Again, the $\Theta$-function in eq.~\eqref{kernel_IC_gamma} ensures energy conservation upon integration of $E'$ over the range $[E, \infty]$.

The total rate for inverse Compton scattering entering the calculation of the electron and positron spectrum is given by~\cite{Jones:1968zza, Kawasaki:1994sc}\footnote{Correcting a typo in eq.~(48) of~\cite{Kawasaki:1994sc}.}
\begin{align}
\Gamma_{e^\pm}^\text{(IC)}(E) = 2\pi \alpha^2 \times \frac{1}{E^2}\int_{0}^{\infty} \text{d}E_\gamma\; \int_{0}^{\infty} \text{d}\be\; \frac{\fpdi_\gamma(\be)}{\be} F(E_\gamma, E, \be) \eqsp.
\end{align}
Finally, the differential rate for the production of electrons and positrons can be written as~\cite{Jones:1968zza, Kawasaki:1994sc}
\begin{align}
K_{e^\pm \rightarrow e^\pm}^\text{(IC)}(E, E') = 2\pi\alpha^2\times \frac{1}{E'^2} \int_{0}^{\infty} \text{d}\be\;\frac{\fpdi_\gamma(\be)}{\be}\;F(E'+\be - E, E', \be) \eqsp.
\end{align}

\subsection*{Additional processes not considered in our calculation}
Other processes such as
\begin{itemize}
	\item Coulomb scattering $e^\pm e^-_\text{th} \rightarrow e^\pm e^-$ and $N e^-_\text{th} \rightarrow N e^-$,
	\item Thompson scattering $N \gamma_\text{th} \rightarrow N \gamma$,
	\item Magnetic moment scattering $N e^-_\text{th} \rightarrow N e^-$ or
	\item Electron-positron annihilation $e^+ e^-_\text{th} \rightarrow \gamma \gamma$
\end{itemize}
are suppressed by the small density of background electrons or nuclei  $\eq{n}_{e}, n_N \ll \eq{n}_{\gamma}$ and can therefore be neglected.

\section{Some benchmarks}
\label{sec:benchmarks_single}

In this appendix we present benchmarks in order to better understand the actual runtime of the code. We ran the script \texttt{decay} with different values of $m_\phi$ and $\tau_\phi$ while fixing $n_\phi/n_\gamma|_{T_0} = 10^{-10}$ at $T_0=10\,\mathrm{MeV}$, $\text{BR}_{ee} = 0$ and $\text{BR}_{\gamma\gamma}=1$ (changing the latter parameters does not change the runtime). Using one core of an \texttt{Intel Core i5-6500 CPU @ 3.20GHz}, we obtain the following results
\begin{center}
	\texttt{./decay $m_\phi\;\mathrm{[MeV]}$  $\tau_\phi\;\mathrm{[s]}$ 10 1e-10 0 1}
	\begin{tabular}{|c|c|c|c|}
		\hline
		$m_\phi\;\mathrm{[MeV]}$ & $\tau_\phi\;\mathrm{[s]}$ &  runtime (with db) & runtime (without db) \\
		\hline\hline
		$10$ & $10^5$ & 38s & 7min 35s\\
		\hline
		$50$ & $10^5$ & 1min 56s & 34min 17s \\
		\hline
		$100$ & $10^5$ & 2min 35s & 55min 44s \\
		\hline\hline
		$10$ & $10^7$ & 24s & 3min 30s\\
		\hline
		$50$ & $10^7$ & 1min 45s & 19min 16s\\
		\hline
		$100$ & $10^7$ & 2min 26s & 32min 23s\\
		\hline
	\end{tabular}
\end{center}
Given these results it is clear that the usage of the database files is highly recommended as it speeds up the calculation by up to a factor of 20. It is also worth noting that the runtime critically depends on the values of $m_\phi$ and $\tau_\phi$, since these parameters determine the energy and temperature range that is used for the calculation. For the \texttt{decay} model the relevant intervals are given by $[E_\text{min}, m_\phi/2]$ and $[10^{1/2}T(\tau_\phi), 10^{-3/2}T(\tau_\phi)]$, respectively. Since the number of points per decade are fixed per default, larger values of $m_\phi$ lead to a larger energy grid and thus to a longer runtime.

Benchmarking the annihilation model we ran the script \texttt{annihilation} for different values of $m_\chi$ while fixing $a=10^{-25}\;\mathrm{cm^3/s}$, $b=0$, $T_\text{kd} = 0$, $\text{BR}_{ee} = 0$ and $\text{BR}_{\gamma\gamma}=1$ (changing the latter parameters again does not change the runtime). In this case, by using the same CPU, we find
\begin{center}
	\texttt{./annihilation $m_\chi\;\mathrm{[MeV]}$ 1e-25 0 0 0 1}
	\begin{tabular}{|c|c|c|c|}
		\hline
		$m_\phi\;\mathrm{[MeV]}$ &  runtime (with db) & runtime (without db) \\
		\hline\hline
		$10$ & 3min 25s & 23min 12s\\
		\hline
		$50$ & 10min 4s & 80min 2s \\
		\hline
		$100$ & 14min 15s & 117min 33s \\
		\hline
	\end{tabular}
\end{center}
Again we find that the runtime is increased for larger values of $m_\chi$, which determines the energy range $[E_\text{min}, m_\chi]$.

We thus conclude that, depending on the model and choice of parameters, the runtime (with database files) can easily vary between $\mathcal{O}(10\,\mathrm{s})$ and $\mathcal{O}(10\,\mathrm{min})$. In order to still enable efficient (and fast) parameter scans, \acropolis comes with a dedicated scanning framework, cf.\ section~\ref{sec:scans}.

\bibliography{references}
\bibliographystyle{JHEP}

\end{document}